\documentclass[aps,nofootinbib,superscriptaddress,floatfix,twocolumn,pra]{revtex4-1}
\usepackage{graphicx}
\usepackage{epstopdf}

\usepackage{amsmath,amsthm,amssymb,latexsym,amsfonts,times,mathrsfs,verbatim}
\usepackage{bm}
\usepackage{mathptmx}
\usepackage{braket}

\DeclareMathAlphabet{\mathcal}{OMS}{cmsy}{m}{n}

\begin{document}

\title{Two-level system damping in a quasi-one-dimensional optomechanical resonator} 

\author{B.D. Hauer}\email{bhauer@ualberta.ca}
\affiliation{Department of Physics, University of Alberta, Edmonton, Alberta, Canada T6G 2E9}
\author{P.H. Kim}
\affiliation{Department of Physics, University of Alberta, Edmonton, Alberta, Canada T6G 2E9}
\author{C. Doolin}
\affiliation{Department of Physics, University of Alberta, Edmonton, Alberta, Canada T6G 2E9}
\author{F. Souris}
\affiliation{Department of Physics, University of Alberta, Edmonton, Alberta, Canada T6G 2E9}
\affiliation{Institut N\'eel, CNRS and Universit\'e Grenoble Alpes, Grenoble, France 38042}
\author{J.P. Davis}\email{jdavis@ualberta.ca}
\affiliation{Department of Physics, University of Alberta, Edmonton, Alberta, Canada T6G 2E9}

\date{\today}

\begin{abstract}

Nanomechanical resonators have demonstrated great potential for use as versatile tools in a number of emerging quantum technologies. For such applications, the performance of these systems is restricted by the decoherence of their fragile quantum states, necessitating a thorough understanding of their dissipative coupling to the surrounding environment. In bulk amorphous solids, these dissipation channels are dominated at low temperatures by parasitic coupling to intrinsic two-level system (TLS) defects, however, there remains a disconnect between theory and experiment on how this damping manifests in dimensionally-reduced nanomechanical resonators. Here, we present an optomechanically-mediated thermal ringdown technique, which we use to perform simultaneous measurements of the dissipation in four mechanical modes of a cryogenically-cooled silicon nanoresonator, with resonant frequencies ranging from 3 - 19 MHz. Analyzing the device's mechanical damping rate at fridge temperatures between 10 mK - 10 K, we demonstrate quantitative agreement with the standard tunneling model for TLS ensembles confined to one dimension. From these fits, we extract the defect density of states ($P_0 \sim$ 1 - 4 $\times$ 10$^{44}$ J$^{-1}$ m$^{-3}$) and deformation potentials ($\gamma \sim$ 1 - 2 eV), showing that each mechanical mode couples on average to less than a single thermally-active defect at 10 mK.

\end{abstract}

\maketitle

\section{Introduction}
\label{intro}

Over the past decade, a number of quantum phenomena have been observed in nanomechanical resonators, including motional ground state cooling \cite{oconnell_2010, teufel_2011, chan_2011}, preparation into squeezed and entangled states \cite{wollmann_2015,pirkkalainen_2015,riedinger_2018,ockeleon-korppi_2018} and nonclassical interaction with electromagnetic fields \cite{palomaki_2013a, riedinger_2016,reed_2017}. This level of quantum control has generated significant interest for the use of nanoresonators in various quantum applications, such as coherent interfacing between two nonclassical degrees of freedom \cite{hill_2012, bochmann_2013, andrews_2014}, storage of quantum information \cite{oconnell_2010, riedinger_2016, reed_2017} and quantum-limited metrology \cite{teufel_2009, kim_2016}. In each of these applications, it is crucial that the nanomechanical resonator maintain its quantum coherence for the duration of the intended operation. For instance, to perform quantum state transfer between the optical and mechanical degrees of freedom of an optomechanical resonator -- a prerequisite for numerous mechanically-mediated quantum information protocols \cite{stannigel_2012, wang_2012, verhagen_2012, hill_2012, palomaki_2013a, palomaki_2013b, riedinger_2016, reed_2017} -- the phononic and photonic modes of the device must couple to each other faster than the rate at which the phononic state decoheres \cite{aspelmeyer_2014}. For a mechanical resonator with angular frequency, $\omega_{\rm m}$, coupled at its damping rate, $\Gamma$, to an environmental bath at a temperature, $T$, this rate is given by $\Gamma_{\rm th} = n_{\rm th} \Gamma$, where $n_{\rm th} = \left( e^{\hbar \omega_{\rm m} / k_{\rm B} T} - 1 \right)^{-1}$ is the bath's thermal phonon occupation \cite{aspelmeyer_2014}. Therefore, in order to minimize decoherence in nanomechanical resonators, such that they can be used as a viable quantum resource, it is critical to focus on understanding their low temperature damping mechanisms.

Though dissipation in mechanical systems can arise from a number of sources, at cryogenic temperatures energy loss is often caused by coupling between the motion of the resonator and its intrinsic material defects \cite{mohanty_2002}. In the simplest treatment, these defects can be modeled as two-level systems (TLS) with an energy separation $E =\sqrt{\Delta^2 + \Delta_0^2}$, realized by tunneling with a characteristic energy, $\Delta_0$, between the two lowest-energy configurational states of the defect, which are split by an asymmetry energy, $\Delta$ \cite{phillips_1987}. For non-resonant defect-phonon interactions at low frequencies ($\hbar \omega_{\rm m} < E$), local strain variations due to the motion of the resonator distort the environment of the TLS defects, perturbing this energy separation away from its static value. The TLS ensemble will then relax towards this new thermal equilibrium via interactions with surrounding phonons at a rate 
\begin{equation}
\tau^{-1} = \frac{\gamma^2}{\rho_{D} c_{\rm e}^{D+2}} \frac{\pi S_{D-1}}{ (2 \pi)^D \hbar^{D+1} } E^{D-2} \Delta_0^2~{\rm coth} \left( \frac{E}{2 k_{\rm B} T} \right),
\label{tauinv}
\end{equation}
which is strongly dependent on the dimensionality, $D$, of the system \cite{behunin_2016}. Here, $\gamma$ is the deformation potential of the device, which characterizes the coupling between the TLS and the motion of the resonator, while $\rho_D$ and $c_{\rm e}$ are the $D$-dimensional mass density ($\rho_3 = \rho$ is the conventional, three-dimensional density) and effective speed of sound of the resonator's material, with $S_D$ being the surface area of the $D$-dimensional unit hypersphere. This finite relaxation rate introduces a phase lag for phonons that interact with defects in the solid, leading to a TLS-induced damping rate of the form \cite{phillips_1987}
\begin{equation}
\begin{gathered}
\Gamma_{\rm TLS} = \frac{\gamma^2}{\rho c_q^2 k_{\rm B} T} \int_0^\infty \int_0^\infty \left( \frac{\Delta}{E} \right)^2 {\rm sech}^2 \left( \frac{E}{2 k_{\rm B} T} \right) \\
\times \frac{\omega_{\rm m}^2 \tau}{1 + \omega_{\rm m}^2 \tau^2} h(\Delta,\Delta_0) d \Delta d \Delta_0,
\label{GamTLS}
\end{gathered}
\end{equation}
where $c_q$ is the mode-dependent effective speed of sound and we have integrated over the energy distribution of the TLS ensemble, represented by the function $h(\Delta,\Delta_0)$ (see Appendix \ref{STMmodel}). For TLS that are amorphous in nature, $h(\Delta,\Delta_0) = P_0 / \Delta_0$ is assumed, where $P_0$ is a constant that characterizes the density of states of the TLS ensemble, as this choice of $h(\Delta,\Delta_0)$ effectively models the broad distribution in TLS separation energies associated with a disordered environment \cite{phillips_1987}. Using this energy distribution function, along with the relaxation rate of Eq.~\eqref{tauinv}, one finds a low temperature dissipation that obeys $\Gamma_{\rm TLS} \sim T^D$, whereas at high temperatures, the damping rate plateaus to a dimensionally-independent constant \cite{behunin_2016}. Therefore, one must carefully consider the dimensionality of the system in question, which will become reduced if the typical thermal phonon wavelength is longer than one or more of the device's characteristic dimensions \cite{seoanez_2008,behunin_2016}. 

This relaxation damping model has been very successful in describing the absorption of sound waves in bulk amorphous solids, where a $T^3$-dependence in acoustic attenuation has been observed at low temperatures for a number of glassy materials in accordance with their three-dimensional nature \cite{pohl_2002}. However, the situation becomes significantly more complicated when considering the reduced geometries associated with nano/micromechanical resonators. Although a linear temperature dependence in mechanical dissipation was first observed for early cryogenic measurements on cm-scale single-crystal silicon torsional oscillators \cite{kleiman_1987, mihailovich_1990}, this behaviour was rationalized as being due to the crystalline nature of the resonator material \cite{phillips_1988} or electronic defects \cite{keyes_1989}, as opposed to reduced dimensionality effects. While a similar linear trend was later reported in polycrystalline aluminum nanobeams \cite{hoehne_2010}, the vast majority of cryogenic dissipation measurements performed on driven micro/nanomechanical resonators have demonstrated a considerably weaker low temperature dependence of $\Gamma \sim T^{1/3}$ \cite{zolfagharkhani_2005, shim_2007, imboden_2009, huttel_2009}. Attempts to explain this sublinear temperature dependence have associated it with the large strain induced by the external drive fields applied to these resonators \cite{ahn_2003} or possibly their beamlike geometries \cite{jiang_2004, seoanez_2008}, however, a full quantitative description has yet to be found. In light of this disconnect between theory and experiment, a clear and careful analysis of TLS damping in reduced-dimensionality nanomechanical resonators is required in order to elucidate this dissipation mechanism \cite{behunin_2016}.

Here, we present measurements of the dissipation in a thermally-driven silicon nanomechanical resonator using a simplified version of the optomechanically-mediated ringdown technique developed by Meenehan {\it et al.}~\cite{meenehan_2015}. Our method circumvents the need for single photon detectors, as well as the requirement that the device must exist in the sideband-resolved regime, all while measuring the Brownian motion of the device to avoid any effects that may arise from large strains due to an external drive \cite{ahn_2003}. Using this measurement scheme, we are able to determine the mechanical damping rate for four of the device's mechanical modes over three orders of magnitude in fridge temperature, ranging from 10 mK to 10 K. Fitting these data, we demonstrate quantitative agreement with the standard tunneling model for damping due to TLS defects embedded in a one-dimensional geometry. Extracting information about the density of states and deformation potentials of the TLS ensembles that couple to the resonator's motion, we speculate that they are caused by glassy surface defects \cite{seoanez_2008, gao_2008} created during fabrication of the device \cite{lu_2005,borselli_2006}. Finally, we show that at 10 mK each mechanical mode couples on average to less than a single thermally-active defect, entering the regime where quantum-coherent interactions between phonons and an individual defect may be possible \cite{ramos_2013}.

\section{Cryogenic Optomechanical Ringdown Measurements}
\label{cryomeas}

\begin{figure*}[t!]
\centerline{\includegraphics[width=6in]{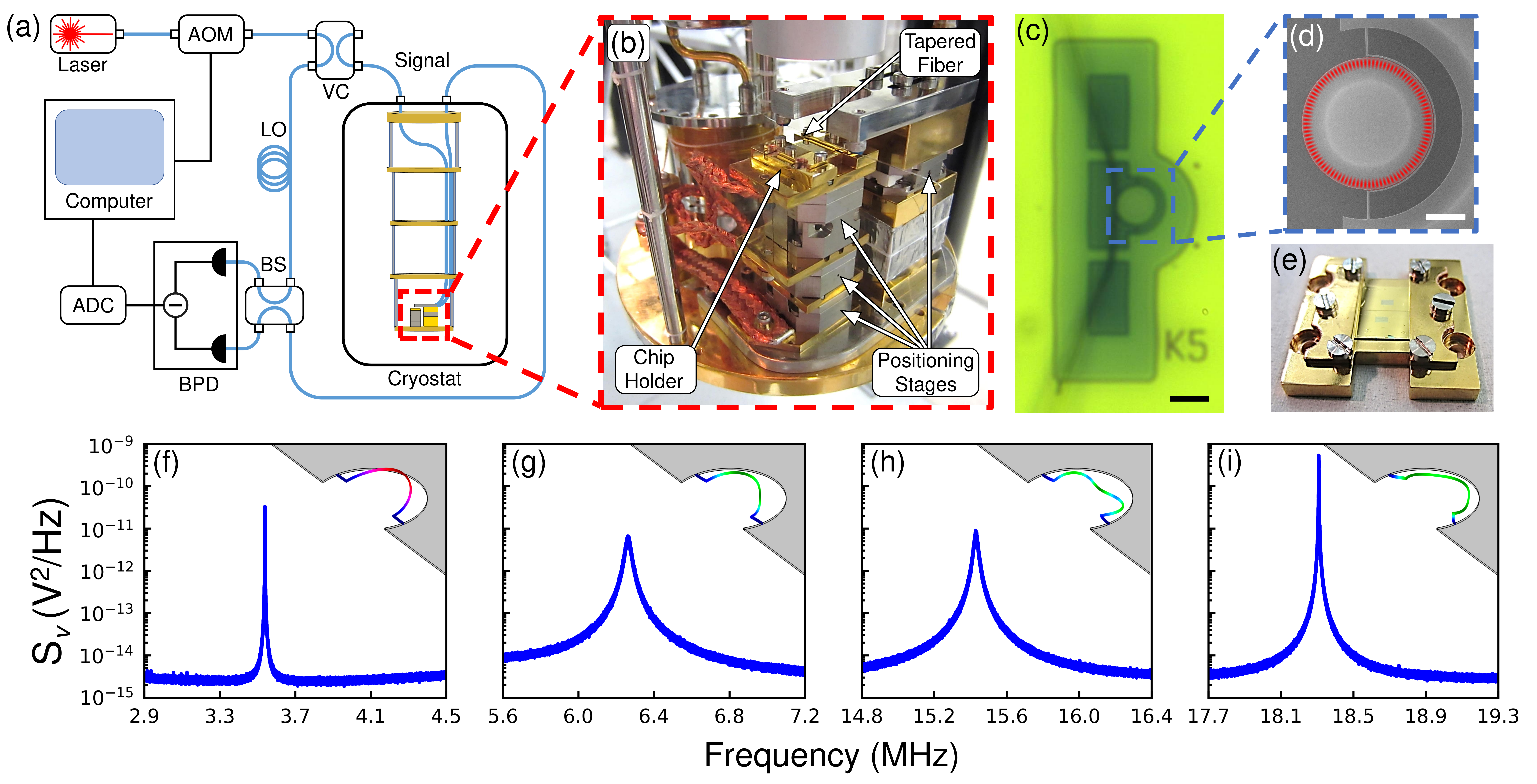}}
\caption{{\label{fig1}} (a) Simplified schematic of the gated homodyne optomechanical detection system (see Appendix \ref{expdet} for details). AOM = acousto-optic modulator, VC = variable coupler, LO = local oscillator, BS = beam splitter, BPD = balanced photodetector, ADC = analog-to-digital converter. (b) Low-temperature coupling apparatus on the mixing chamber plate of the dilution refrigerator (see Ref.~\cite{macdonald_2015} for details). (c) Optical microscope image highlighting the 210 nm thick gold layer (yellow) which is deposited onto the silicon to within $\sim$10 $\mu$m of the device to improve thermalization at low temperatures. (d) A zoomed-in scanning electron micrograph of the optomechanical device studied in this work. Overlaid in red is a finite element method (FEM) simulation showing the magnitude of the electric field for the optical whispering gallery mode used in this work. Scale bars are (c) 10 $\mu$m and (d) 3 $\mu$m. (e) Image of the chip in its gold-plated copper holder. (f)-(i) The voltage spectral density, $S_v$, obtained by continuously monitoring the resonator's mechanical motion in exchange gas at 4.2 K. Measurements were performed with 10 $\mu$W of optical power input to the fridge (corresponding to an input power of $P_{\rm in}$ = 7.5 $\mu$W at the microdisk) and 2.6 mW in the LO. Inset are FEM simulations of the displacement profiles for (f) the fundamental out-of-plane torsional mode (3.53 MHz), along with the (g) ``side-to-side'' (6.28 MHz), (h) antisymmetric ``breathing-like'' (15.44 MHz) and (i) symmetric ``breathing-like'' (18.31 MHz) in-plane flexural modes. Red (green) indicates out-of-plane (in-plane) motion, while blue denotes zero displacement.}
\end{figure*}

The optomechanical device studied in this paper [see Fig.~\ref{fig1}(c),(d)] consists of a half-ring mechanical resonator (width $w =$ 200 nm, thickness $d =$ 250 nm) partially surrounding a whispering gallery mode microdisk cavity (see Appendix \ref{optoprop} for details), both of which are fabricated from single-crystal silicon. The device chip is thermally anchored to the mixing chamber plate of a dilution refrigerator and measured using a gated homodyne detection scheme [see Fig.~\ref{fig1}(a)], capable of simultaneously transducing the motion of several mechanical modes of the half-ring resonator [Fig.~\ref{fig1}(f)-(i)] with sub-microsecond resolution in the time domain.

To perform these measurements, we excite the first order radial mode of the optomechanical cavity [azimuthal mode number $M =$ 49 -- see Fig.~\ref{fig1}(d)], with resonant frequency $\omega_{\rm c}/ 2 \pi =$ 188.8 THz ($\lambda_{\rm c} =$ 1587.9 nm) and linewidth $\kappa / 2 \pi =$ 1.0 GHz ($Q_{\rm o} =$ 1.9$\times$10$^5$). Due to the large disparity between the energy of this optical mode (hundreds of THz) and the mechanical modes (tens of MHz), coupled with the diminishing thermal conductivity of silicon at low temperatures \cite{holland_1963}, even small input powers to the optical cavity act to rapidly heat the mechanics. This heating effect can be modeled by considering a mechanical mode simultaneously coupled at its intrinsic damping rate, $\Gamma_{\rm i}$, to the thermal environment of the fridge, and at a rate, $\Gamma_{\rm p}$, to a hot phonon bath generated by either the absorption of cavity photons or radiation pressure backaction (or a combination of the two), with the total damping rate of the system given by $\Gamma = \Gamma_{\rm i} + \Gamma_{\rm p}$. If light is coupled into the optical cavity at time $t=t_0$, the average phonon occupancy of the mechanical mode as a function of time is then given by \cite{meenehan_2015}
\begin{equation}
\braket{n}(t) = \braket{n}(t_0) e^{- \Gamma (t-t_0)} + n_{\rm eq} \left( 1 - e^{- \Gamma (t-t_0)} \right).
\label{heateq}
\end{equation}
Here, $n_{\rm eq} = \left( n_{\rm th} \Gamma_{\rm i} + n_{\rm p} \Gamma_{\rm p} \right)/\Gamma$ is the equilibrium phonon occupation of the mode, with $n_{\rm th}$ and $n_{\rm p}$ being the average phonon occupancies of the environmental and photon-induced baths, respectively (see Appendix \ref{heatmodel}). We note that for the temperatures considered here, $n_{\rm p} \gg n_{\rm th}$ and $\Gamma \approx \Gamma_{\rm p}$, such that $n_{\rm eq} \approx  n_{\rm p}$.

\begin{figure}[t!]
\centerline{\includegraphics[width=3.0in]{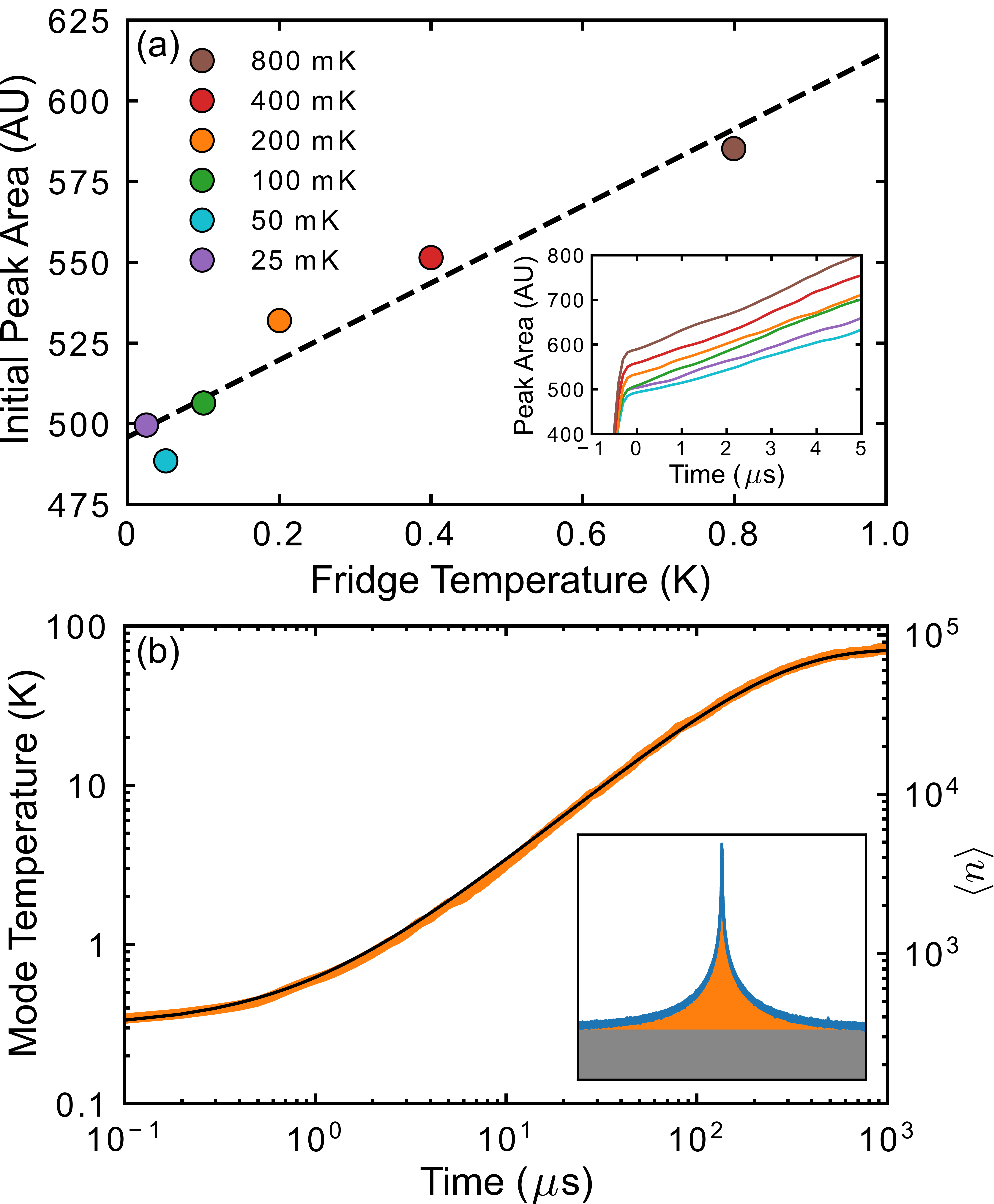}} 
\caption{{\label{fig2}} (a) Area under the peak (including contributions from both the mechanical signal and imprecision noise) at the beginning of the measurement pulse for the 18.31 MHz mode plotted versus fridge temperature. The uncertainty in each point is smaller than the marker size. For the high input power used here ($P_{\rm in}$ = 75 $\mu$W), the shot-noise of the optomechanical measurement is sufficiently suppressed to resolve the mode's initial phonon occupation. Fitting this linear trend (dashed line), we calibrate the peak area in terms of the mechanical mode temperature, with the y-intercept indicating an imprecision noise floor equivalent to 29.5 fm$^2$/Hz (see Appendix \ref{mechT}). Inset highlights the rapid increase in the peak area during the first 5 $\mu$s of the measurement for each temperature (color-coded to match the main figure), taken by averaging data from 5000 individual optical pulses 4 ms in length and scaled by discarding the initial 20 $\mu$s of transient signal due to the applied numerical bandpass filter. Each data point in the main figure is extracted from a fit of Eq.~\eqref{heateq} over the full pulse duration of this data. The rapid settling of the signal for $t<0$ is set by the 10 MS/s effective sampling rate of our data acquisition (see Appendix \ref{expdet}). (b) A typical heating curve corresponding to the point in (a) with the fridge temperature at 200 mK. The data (orange) is calibrated in terms of both temperature and average phonon occupancy, showing that the mechanical mode heats to $T \approx 80$ K ($\braket{n} \approx 10^5$) within the first millisecond of the measurement pulse. The solid black line is a fit to Eq.~\eqref{heateq}, used to extract the initial and final phonon occupancy of the mode. Inset shows the continuously-monitored spectral density of the mechanical resonance over the $\sim$1.2 MHz bandwidth window used for these measurements. Orange illustrates the area under the peak due to mechanical motion, while grey indicates the noise floor.}
\end{figure}

\begin{figure}[t!]
\centerline{\includegraphics[width=3.0in]{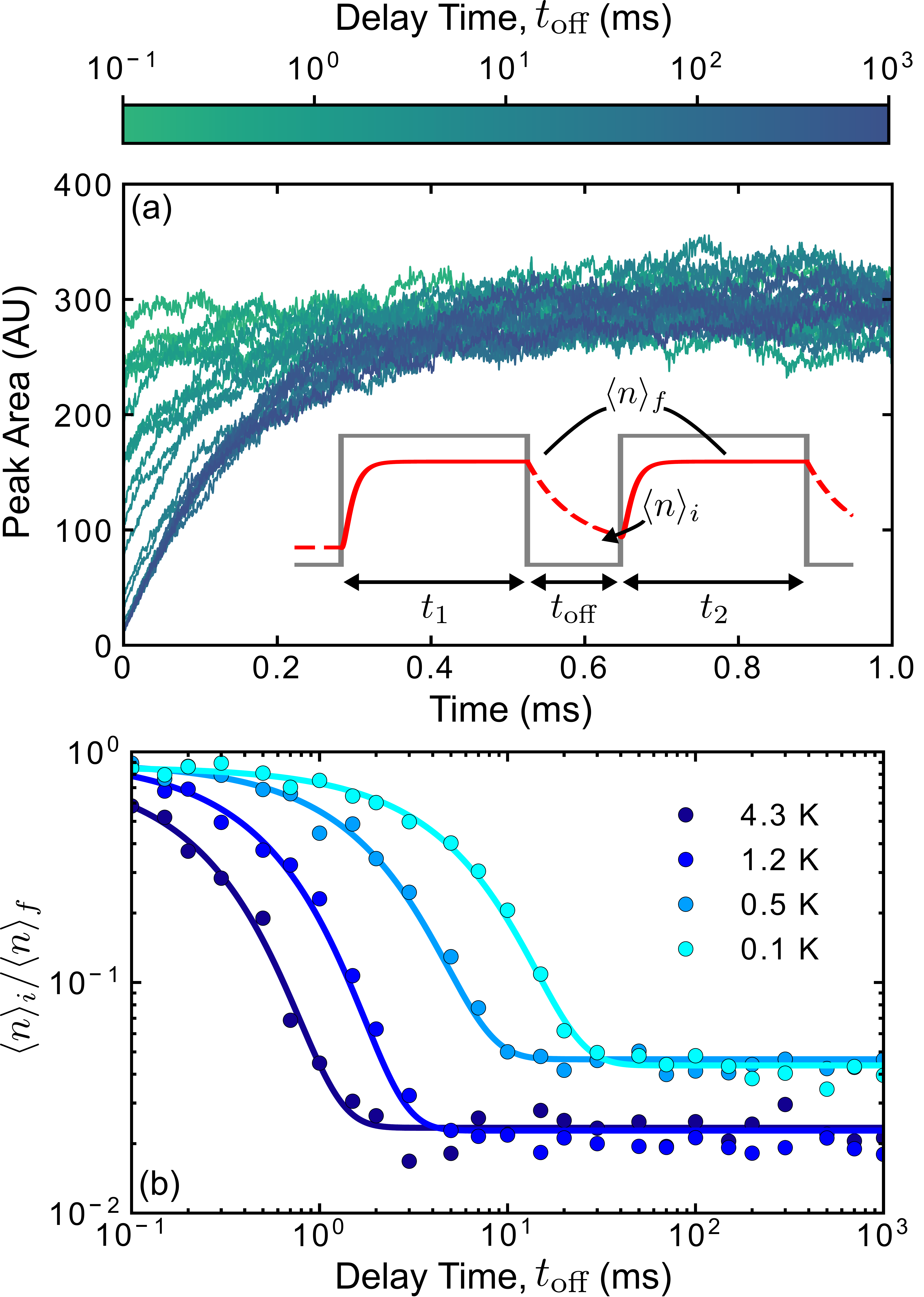}}
\caption{{\label{fig3}} (a) The first millisecond of pulsed data for the 18.31 MHz mechanical mode, obtained by averaging 500 individual probe pulses with the fridge at its base temperature of 10 mK. Measurements are performed by varying the delay time between pump and probe pulses (each with $P_{\rm in}$ = 7.5 $\mu$W and a full duration of 2 ms) as indicated by the color bar. These traces are fit with Eq.~\eqref{heateq} to extract their initial and final occupations. Inset is a schematic of the pump-probe sequence. The grey line indicates the state of the laser (high = on, low = off), with the solid (dashed) red line being the average occupancy of the mechanical mode with the laser on (off). (b) Thermally excited ringdown measurements at a number of fridge temperatures. The solid lines are fits to the data using Eq.~\eqref{cooleq}, allowing for extraction of the intrinsic damping rate, $\Gamma_{\rm i}$, at each temperature. The disparity in the noise floor between the low temperature (0.1 K, 0.5 K) and high temperature (1.2 K, 4.3 K) data results from varying levels of optomechanical transduction ({\it i.e.}, small variations in input power, optical coupling to the device, etc.)~between data runs and does not have an effect on the extracted intrinsic mechanical damping rate.}
\end{figure}

This rapid heating (see Fig.~\ref{fig2}) prevents one from continuously monitoring the device's motion at low temperatures. However, by performing time-resolved measurements of the mechanical resonator (see Appendix \ref{expdet}) and looking at its phonon occupancy for $t < 1$ $\mu$s, we show that the device is initially thermalized to the fridge, reaching an average phonon occupancy as low as $\braket{n} \approx 28$ for the 18.31 MHz mode at $T$ = 25 mK and allowing for complete calibration of the device temperature at all times during the measurement (see Fig.~\ref{fig2}). Furthermore, we capitalize on this optically-induced heating to implement a pump/probe measurement technique \cite{meenehan_2015}, as illustrated in the inset of Fig.~\ref{fig3}(a). This allows us to observe the thermalization of the laser-heated mechanical mode back to the fridge temperature at its intrinsic damping rate according to
\begin{equation}
\frac{\braket{n}_i}{\braket{n}_f} = \frac{\left(n_{\rm eq} - n_{\rm th} \right) e^{-\Gamma_{\rm i} t_{\rm off}} + n_{\rm th} + n_{\rm imp}}{n_{\rm eq} + n_{\rm imp}}.
\label{cooleq}
\end{equation}
Here, $\braket{n}_i$ and $\braket{n}_f$ are the measured phonon occupancies of the mechanical mode (including the apparent contribution, $n_{\rm imp}$, due to imprecision noise) at the beginning of the probe pulse and at the end of the pump pulse, respectively, while $t_{\rm off}$ is the time delay between turning off the pump pulse and turning on the probe pulse (see Appendix \ref{heatmodel}). In Eq.~\eqref{cooleq}, as well as the experiment, we have chosen the lengths of the pump pulse, $t_1$, and probe pulse, $t_2$, to be equal, as well as satisfy $t_1 = t_2 \gg \Gamma^{-1}$ such that $\braket{n}_f = n_{\rm eq} + n_{\rm imp}$ at the end of each pulse. By varying the delay between pulses and fitting the data to Eq.~\eqref{cooleq}, as seen in Fig.~\ref{fig3}(b), we can extract the intrinsic mechanical damping rate of the device, allowing us to map out its low-temperature dependence. We note that in order to achieve sub-microsecond time resolution for our measurements, we must integrate the mechanical spectra over a relatively large bandwidth of $\sim$1.2 MHz. This prevents us from measuring TLS-induced resonance frequency shifts, as has been previously observed in other mechanical systems \cite{kleiman_1987, mihailovich_1990, zolfagharkhani_2005, shim_2007, imboden_2009, huttel_2009}, as well as superconducting microwave circuits \cite{gao_2008, suh_2012, suh_2013}. However, if one were to reduce this bandwidth, at the expense of time resolution, it may be possible to track the mechanical resonance frequency of the device as it heats up during measurement.

\section{Quantitative Agreement with the One-Dimensional Standard Tunneling Model}
\label{quantagree}

\begin{figure*}[t!]
\centerline{\includegraphics[width=6in]{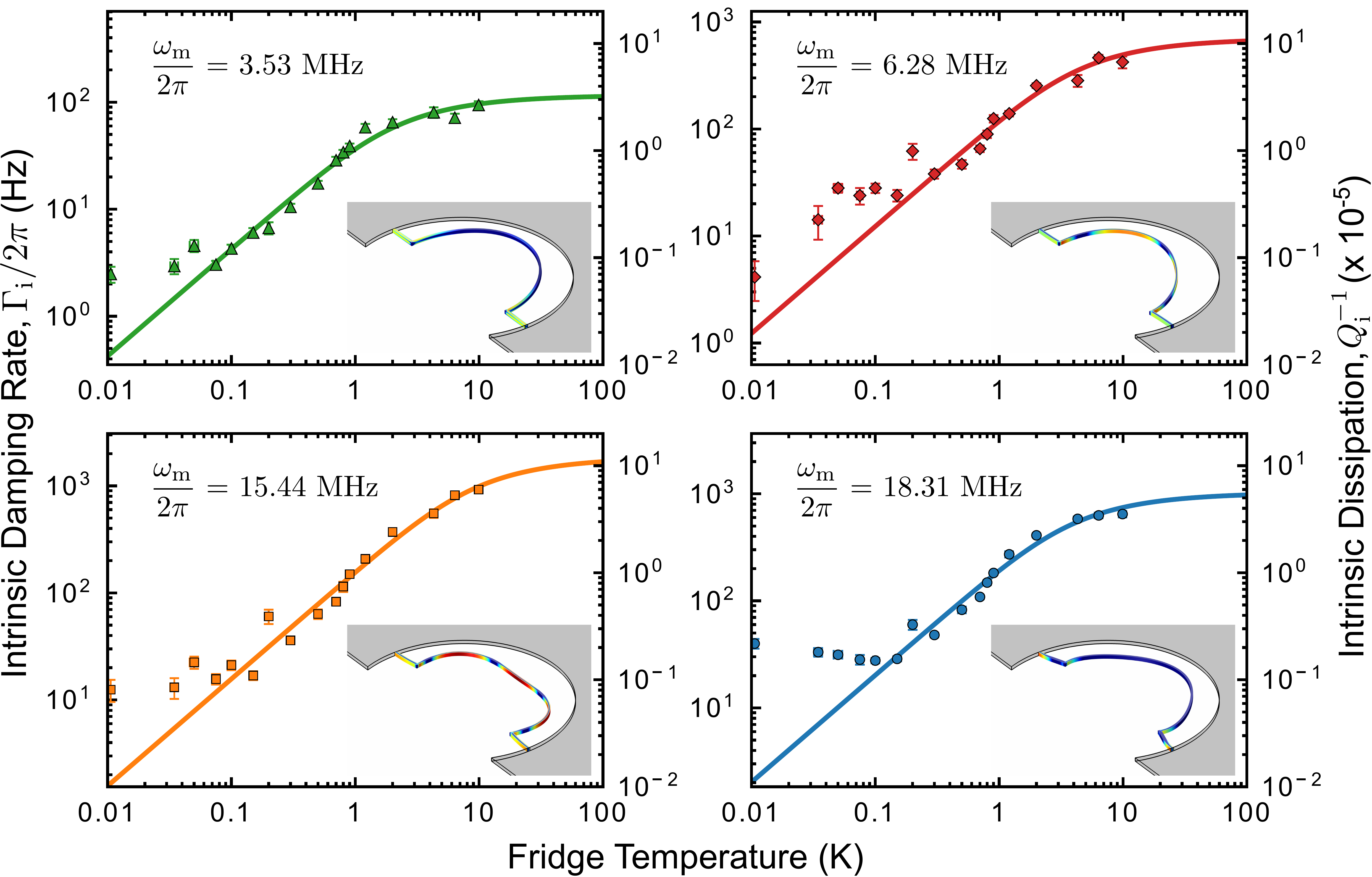}}
\caption{{\label{fig4}} The intrinsic damping rate, $\Gamma_{\rm i}$, measured for each of the four studied mechanical modes plotted versus fridge temperature, with the right axis displaying their intrinsic dissipation, $Q_{\rm i}^{-1} = \Gamma_{\rm i}/\omega_{\rm m}$. Markers in each plot represent the experimentally-determined damping rate extracted from fits of Eq.~\eqref{cooleq} to data similar to that seen in Fig.~\ref{fig3}(b), with error bars representing a single standard deviation in the uncertainty of the fit. Solid lines are fits to Eq.~\eqref{GamTLS}, using $D=1$ in Eq.~\eqref{tauinv} and $h(\Delta,\Delta_0) = P_0 / \Delta_0$, demonstrating the temperature dependence of the mechanical damping rate according to a one-dimensional TLS interaction model. From the fits, we obtain the parameters $P_0$ and $\gamma$ for the TLS ensemble coupled to each mechanical mode, which are given in Table \ref{TLStab}. For $T \lesssim$ 100 mK, the damping rate plateaus to a relatively constant value, which could be due to a number of effects, including temperature independent radiation of acoustic energy into the substrate \cite{cross_2001, wilsonrae_2008} or heating of the chip due to measurement (see Appendix \ref{mechT}). Inset are the logarithm of the normalized strain energy density simulated using FEM analysis for each mechanical mode. These simulations highlight the fact that the 3.53 MHz and 18.31 MHz modes have smaller spatial strain profiles that are localized to the supports of the half-ring (as characterized by their effective strain volumes - see Table \ref{TLStab}), whereas the strain energy density profiles of the 6.28 MHz and 15.44 MHz modes extend into the rounded portion of the resonator.}
\end{figure*}

Measurements of the damping rate for each of the four studied mechanical modes are performed with fridge temperatures varying from 10 mK to 10 K. While each mode exhibits qualitatively similar behaviour, as seen in Fig.~\ref{fig4}, in order to quantitatively analyze the data, we must first determine the dimensionality of the resonator. This is done by comparing the transverse dimensions of our device ($w =$ 200 nm, $d =$ 250 nm) to the shortest thermal phonon wavelength present in the system, given by $\lambda_{\rm th} = 2 \pi \hbar c_{t_1}/k_{\rm B} T \approx 225~{\rm nm} \cdot {\rm K}/T$, where $c_{t_1} =$ 4679 m/s is the slowest speed of sound in single-crystal silicon (see Appendix \ref{STMmodel}). Our device therefore behaves one-dimensionally for $T \lesssim 1$ K, however, to simplify the analysis we consider our device to be quasi-one-dimensional for all temperatures considered here. This approximation is justified by the fact that at high temperatures, the TLS-induced damping rate plateaus to a constant value that is independent of the dimensionality of the system \cite{behunin_2016}. Using this approximation, we fit the data in Fig.~\ref{fig4} using the one-dimensional relaxation TLS damping model, found by numerically integrating Eq.~\eqref{GamTLS} while taking $D=1$ for the TLS relaxation rate in Eq.~\eqref{tauinv}. Parameters extracted from these fits are summarized in Table \ref{TLStab}.

\begin{table}[h!]
\begin{tabular}{ ccccc }
\hline
$\omega_{\rm m}/2 \pi$ & $V_{\rm str}$ & $P_0$ & $\gamma$ & $N_{\rm th}$ \\  
 ~(MHz)~ & ~($\times 10^{-21}$ m$^3$)~ & ~(J$^{-1}$ m$^{-3}$)~ & ~(eV)~ & @ 10 mK \\ \hline
\hline
3.53 & 3.6 (0.44\%) & 9.7 $\times 10^{43}$ & 1.3 & 0.05 \\

6.28 & 6.3 (0.77\%) & 4.0 $\times 10^{44}$ & 1.2 & 0.35 \\

15.44 & 11 (1.4\%) & 3.6 $\times 10^{44}$ & 1.3 & 0.55 \\

18.31 & 1.7 (0.21\%) & 7.0 $\times 10^{43}$ & 2.2 & 0.02 \\

\hline
\end{tabular}
\caption{Summary of the density of states parameter, $P_0$, and deformation potential, $\gamma$, extracted from fits of the data in Fig.~\ref{fig4} to the one-dimensional TLS relaxation damping model given by Eq.~\eqref{GamTLS}. The uncertainty in each of these fit parameters can be found in Table \ref{TLStaba} of Appendix \ref{tlsfits}. Also included is the number of thermally-active defects, $N_{\rm th} \sim P_0 V_{\rm str} k_{\rm B} T$, calculated using the fridge base temperature of 10 mK for the TLS ensembles contained within the effective strain volume, $V_{\rm str}$, of a given mechanical mode. Each effective strain volume is determined using FEM simulations of the mechanical mode's strain energy density (see Appendix \ref{strain}). The percentage of the total geometric volume, $V =$ 8.1 $\times$ 10$^{-19}$ m$^3$, occupied by each effective strain volume is given in parentheses. As one can see, the two modes with a larger effective strain volume (6.28 MHz, 15.44 MHz) couple to TLS ensembles with a defect density of states approximately four times larger than that observed for the two modes with a smaller effective strain volume (3.53 MHz, 18.31 MHz).}
\label{TLStab}
\end{table}

Upon inspection of these fit parameters, one can immediately see that the 6.28 MHz and 15.44 MHz mechanical modes couple to a defect density ($P_0 \sim$ 4 $\times$ 10$^{44}$ J$^{-1}$ m$^{-3}$) that is approximately four times larger than that sampled by the 3.53 MHz and 18.31 MHz modes ($P_0 \sim$ 1 $\times$ 10$^{44}$ J$^{-1}$ m$^{-3}$). We attribute this disparity in TLS ensemble densities to the fact that these first two modes have a larger extent to their strain energy distribution than the latter two modes (see Fig.~\ref{fig4} inset), as quantified by their effective strain volumes, $V_{\rm str}$ (see Appendix \ref{strain}). This effect could also be enhanced by the fact that the two modes with larger strain volumes have a significant portion of their strain energy density localized to the rounded portion of the ring, where multiple crystal axis orientations are sampled. We also point out that the extracted TLS density parameters are on the order of that observed in bulk amorphous silica \cite{golding_1976a, golding_1976b, golding_1976c}, much larger than what would be expected for crystalline silicon resonators, where the TLS density of states has been found to be at least an order of magnitude smaller \cite{kleiman_1987,phillips_1988}. This is likely due to the significantly larger surface-to-volume ratio of our nanoscale devices, which results in defects at the surface of the resonator \cite{seoanez_2008, gao_2008, lu_2005, borselli_2006} providing a larger contribution to the overall defect density, as has been previously reported in optomechanically-measured gallium arsenide microdisks \cite{hamoumi_2018}. We note that this hypothesis is further supported by the fact that over half of the strain energy for each mechanical mode exists within the first 20 nm of the resonator's surface (see Appendix \ref{strain}).

From $P_0$, we can also determine the total number of thermally-active defects located within the effective strain volume of the resonator as $N_{\rm th} \sim P_0 V_{\rm str} k_{\rm B} T$ \cite{behunin_2016, seoanez_2008}. As can be seen from Table \ref{TLStab}, at the lowest achievable temperature of our fridge (10 mK), the resonator is already at the point of coupling to less than a single defect on average. In this situation, known as the {\it small mode volume limit}, the TLS no longer act as a bath and a fully quantum mechanical description must be applied, resulting in the defect-phonon system undergoing Rabi oscillation \cite{behunin_2016}. It is possible that this is the cause of the mechanical damping rate flattening out to a constant value for $T \lesssim 100$ mK, as in this regime other loss mechanisms, such as radiation of acoustic energy at the resonator's clamping points \cite{cross_2001, wilsonrae_2008}, begin to dominate. An alternative explanation is that this plateau is due to measurement-induced heating of the chip at low temperatures (see Appendix \ref{mechT}).

Finally, the extracted deformation potentials are on the order of $\gamma \sim$ 1$-$2 eV, comparable to the those found in bulk amorphous silica \cite{golding_1976a, golding_1976b}. We point out that these values are notably less than the 3 eV that has been previously reported for TLS defects caused by boron dopants in crystalline silicon \cite{mihailovich_1992}, further supporting the hypothesis that these TLS are caused by glassy defects at the surface of the resonator \cite{seoanez_2008}.

\section{Conclusion}
\label{conc}

We have performed simultaneous optomechanical ringdown measurements of thermally-driven motion for four mechanical modes, with resonant frequencies ranging from 3 to 19 MHz, in a single-crystal silicon nanomechanical resonator. From these low-strain measurements, we extract the damping rate for each mechanical mode over a fridge temperatures ranging from 10 mK to 10 K. Fitting these data to a one-dimensional TLS damping model, we demonstrate that dimensionality has a strong effect on the defect-phonon interaction, which is especially important for the reduced geometries associated with nanoscale resonators. Extracting information about the density of states and deformation potentials of the TLS ensembles, we find that they are consistent with glassy surface defects created during fabrication of the nanoresonator, with a concentration similar to that observed in bulk amorphous silica. Comparing the density of states for the TLS ensembles coupled to each mechanical mode, we find that the two modes exhibiting a larger spatial extent to their strain profiles couple to TLS ensembles roughly four times more dense than those coupling to modes with smaller effective strain volumes. To identify and eliminate these sources of TLS dissipation, one could apply more sophisticated silicon surface treatments, such as passivation and reconstruction in a hydrogen atmosphere \cite{bender_1994}, to reduce defects at the device's surface or use higher resistivity silicon to remove any effects dopants may have \cite{mihailovich_1992}.

At the fridge base temperature of 10 mK we further find that the small effective mode volumes of our device should allow us to achieve coupling to less than an individual thermally-active defect on average for each of the four studied mechanical modes. Defect-phonon coupling on this level opens the door to proposed cavity QED-like experiments between an individual defect and phonons within the resonator, providing a nonlinear quantum interaction which could be used for the storage of quantum information \cite{neeley_2008}, quantum control of a single defect center \cite{arcizet_2011,golter_2016} or nonclassical state preparation of the mechanical element \cite{ramos_2013}. Furthermore, by tailoring the phononic structure and mode frequencies of a nanoresonator, it may be possible to engineer a Purcell-like defect-phonon interaction, leading to enhancement or suppression of TLS radiation into a specific mechanical mode \cite{behunin_2016}. Conversely, one could imagine using the mechanical resonator as a probe of the dynamics of a single quantum defect, furthering our incomplete knowledge of the microscopic nature of TLS defects, as well as their interactions with each other \cite{leggett_1987,fu_1989,agarwal_2013}.

\begin{acknowledgments}

This work was supported by the University of Alberta, Faculty of Science; the Natural Sciences and Engineering Research Council, Canada (Grants Nos. RGPIN-2016-04523, DAS492947-2016, and STPGP 493807 - 16); and the Canada Foundation for Innovation. B.D.H. acknowledges support from the Killam Trusts.

\end{acknowledgments}

\begin{appendix}

\section{Experimental Details}
\label{expdet}

\subsection{Device Fabrication}
\label{devfab}

To fabricate our optomechanical devices, we started with a $\langle100\rangle$ p-doped (boron, 22.5 $\Omega \cdot$cm) silicon-on-insulator (SOI) wafer, consisting of a 250 nm-thick device layer of monocrystalline silicon on top of a 3 $\mu$m-thick sacrificial layer of silicon dioxide supported by a 0.5 mm-thick silicon handle. The wafer was initially diced into 10 mm $\times$ 5 mm chips and cleaned using a hot piranha solution (75\% H$_2$SO$_4$, 25\% H$_2$O$_2$) for 20 min. A masking layer (positive resist, ZEP-520a) was deposited onto the clean silicon device layer to pattern the half-ring/optical disk structure using a 30 kV e-beam lithography system (RAITH150 Two), followed by a cold development at --15 $^\circ$C (ZED-N50). The chip was then reactive-ion etched (C$_4$F$_8$ and SF$_6$) to transfer the pattern to the silicon and subsequently cleaned with piranha so that it could be spun with a new mask (positive photoresist, HPR 504). After optical lithography, Cr and Au layers (7 nm and 210 nm, respectively) were sputtered on both sides of the chip with equal thickness, surrounding the devices with a gold thermalization layer, as shown in Fig.~\ref{fig1}(c). Ultrasonic lift-off in acetone and room-temperature piranha cleaning were then used to ensure the cleanliness of these processed chips. Finally, the chips were immersed in HF solution (49\% HF) for 1 minute to etch the sacrificial oxide layer, as well as passivate the exposed silicon surfaces of our devices \cite{higashi_1990}, which was followed by critical point drying to avoid stiction. We note that through more sophisticated treatment techniques, such as passivation and reconstruction of the silicon surfaces in a hydrogen atmosphere \cite{bender_1994}, it may be possible to reduce the defect density at the surface of the resonator, in turn leading to a reduction in TLS-induced mechanical damping.

\subsection{Gated Homodyne Measurement}
\label{optosetup}

\begin{figure}[h!]
\centerline{\includegraphics[width=\columnwidth]{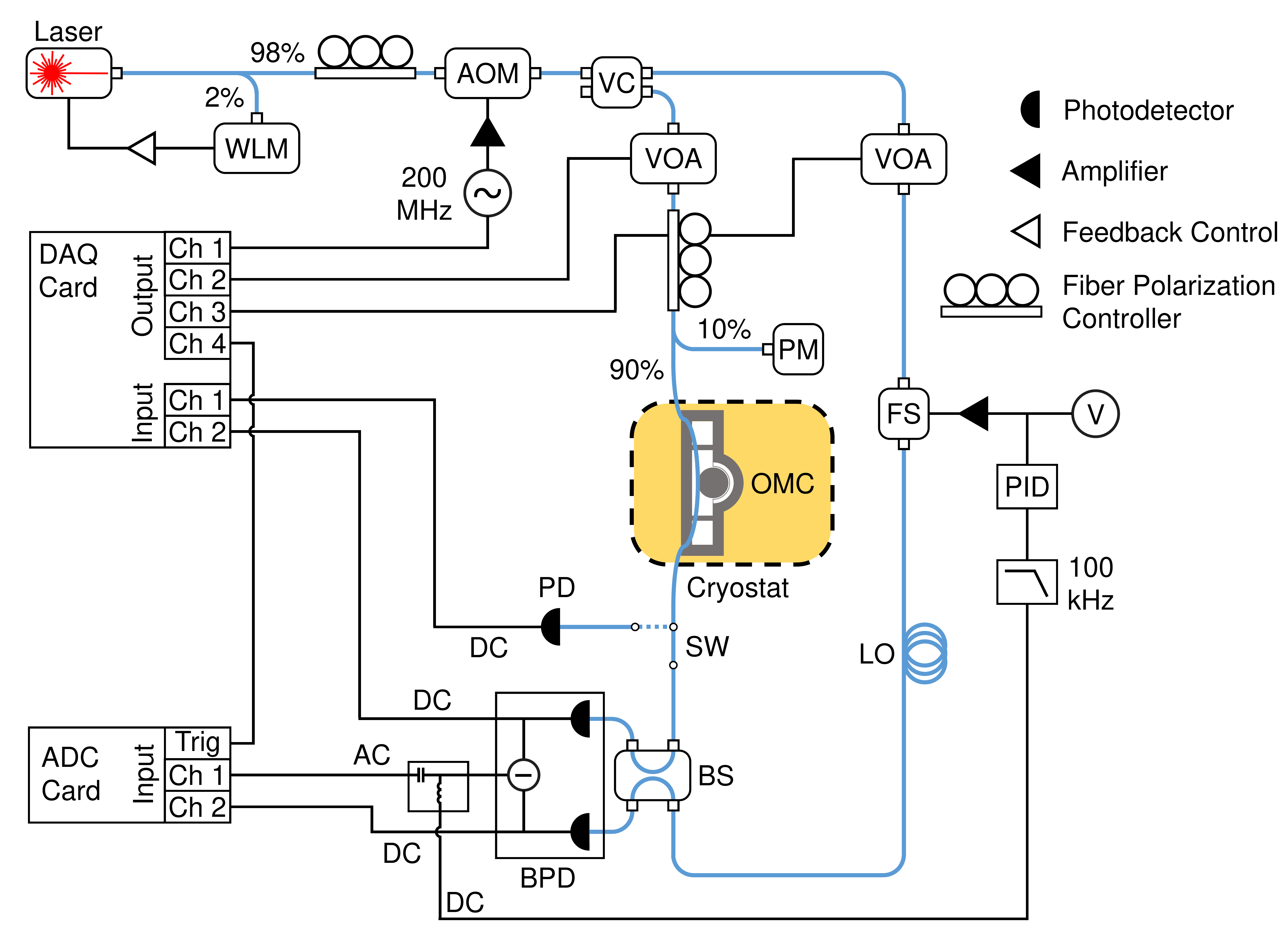}}
\caption{{\label{fig5}} Detailed schematic of the gated homodyne detection system used to perform measurements of the cryogenic optomechanical device presented in this work. WLM = wavelength meter, AOM = acousto-optic modulator, VC = variable coupler, VOA = variable optical attenuator, PM = power meter, OMC = optomechanical cavity, FS = fiber stretcher, PID = proportional-integral-derivative controller, V = voltmeter, LO = local oscillator, SW = optical switch, BS = beam splitter, BPD = balanced photodetector, PD = photodetector, ADC = analog-to-digital converter, DAQ = data acquisition.}
\end{figure}

To measure the motion of our optomechanical device, we implemented a gated optical homodyne detection scheme, a detailed schematic of which can be seen in Fig.~\ref{fig5}. Light from a tunable external cavity diode laser is fiber coupled into the optical circuit, where its wavelength is monitored using a 2\% pick-off to a wavelength meter (WLM), with this reading fed back into the laser controller to ensure long-term frequency stability. The remainder of the signal is sent through an acousto-optic modulator (AOM), allowing for gating of the optical signal with a rise/fall time of $\sim$5 ns, faster than all other timescales associated with the system. The laser light is then sent through a variable coupler, where it is split into two separate beams: the signal and the local oscillator (LO), with the power in each arm set by a voltage-controlled variable optical attenuator (VOA). For the measurements detailed in this work, the LO is kept at a constant power of 2.6 mW, while the power in the signal arm is varied depending on the experiment. The light in the signal arm is coupled into and out of the dilution unit using optical fiber feedthroughs, with its polarization optimized using a fiber polarization controller (FPC) and its power monitored by a power meter (PM). Inside the fridge, a low-temperature dimpled tapered fiber \cite{michael_2007, hauer_2014} is used to inject light into the optomechanical device, while also collecting the optical signal exiting the cavity. After coupling out of the fridge, this optical signal is recombined with the LO via a 50/50 fiber beam splitter (BS), with both outputs sent to a balanced photodetector (BPD). The path length difference between the LO and signal arm of the circuit is maintained by feeding the DC voltage difference signal of the BPD through a proportional-integral-derivative (PID) controller and into a fiber stretcher (FS) located in the LO arm, such that deviations from the optical path length setpoint are compensated for. This process locks the phase of the homodyne measurement and allows for probing of a specific quadrature of the optical field, with the mechanical motion extracted as fluctuations in the AC portion of the BPD's voltage difference signal, which is recorded in the time domain using a 500 MS/s analog-to-digital converter (ADC). The DC voltage readouts from each of the BPD's individual photodetectors are also collected, with one output sent to a low-frequency data acquisition (DAQ) card to monitor slow drifts, while the other is sent to the ADC to observe rapid transients in this signal. Finally, we note that we have included a voltage-controlled optical switch (SW) after the fiber output from the fridge, such that we can opt to toggle the optical signal out of the homodyne loop to a standard, single channel photodetector (PD), allowing for DC spectroscopic measurements of the optical cavity's lineshape.

To perform the pulsed measurements used to measure the mechanical dissipation of our devices, the optomechanical detection system is initially set up by sending a continuous-wave laser signal through the optical circuit. The dimpled tapered fiber is then carefully aligned to couple with the microdisk, after which the laser wavelength is tuned onto resonance with one of the cavity's optical modes and the transduction of the mechanical signal is optimized. We note that due to the relatively high optical powers (10 -- 100 $\mu$W) input to the fridge during this initial set up, the base plate, along with the optomechanical device, heats up significantly. Therefore, once we have ensured that the fiber is in place, the optical circuit is toggled into the ``off'' state by closing the AOM (extinction ratio of 50 dB), preventing optical power from reaching the dimple. After approximately 1 -- 2 hours in this state, the fridge returns to its set-point temperature and is ready for pulsing measurements.

For the double pulse measurement outlined in the inset of Fig.~\ref{fig3}(a), we begin by sending a trigger signal from the DAQ card to a 200 MHz frequency source, activating an output signal that is amplified to 10 V$_{\rm RMS}$ and sent to the AOM. This electrical signal opens the AOM, generating the initial pump pulse that is used to thermally excite the motion of the mechanical resonator. The AOM is left open until the predetermined pulse time, $t_1$, has passed, at which point it is closed by turning the frequency source off with a second signal from the DAQ card. The mechanical resonator is then left in the dark to decay towards thermal equilibrium for a set wait time, $t_{\rm off}$, after which a probe pulse, created in an identical manner to the pump pulse, is sent to access the device. To ensure the data from the probe pulse is recorded, the ADC is activated using another trigger signal generated by the DAQ card at a time chosen to be 10 or 100 $\mu$s -- depending on the length of $t_{\rm off}$ -- before the probe pulse is created. Finally, the AOM is closed after a time, $t_2$, has elapsed following the generation of the probe pulse, returning the optical circuit back to its ``off'' state. Note that for the experiments performed here, we always take $t_1 = t_2$, such that the phonon occupation of the mechanical mode at the end of the pump pulse can be inferred from observation of the probe pulse (see Appendix \ref{heatmodel}), minimizing the amount of data that needs to be collected. After a 200 ms wait to reinitialize the ADC, this procedure is repeated until the desired number of pulses is acquired. Single pulse measurements are performed identically to the double pulse measurements, with the omission of the pump pulse. We note that the gating of the optical circuit is completely controlled by outputs from the DAQ card, ensuring consistent timing referenced to its 1 MHz internal clock.

\subsection{Signal Processing}
\label{sigproc}

The displacement of the half-ring resonator is dispersively coupled to the monitored optical mode, modulating its effective index of refraction, and therefore resonance frequency, such that the mechanical motion is encoded into the fluctuating phase of the optical signal that is transmitted through the cavity. Therefore, once this signal beam is recombined with the LO and sent to the BPD, the mechanical motion is transduced into a time-varying voltage signal, $v(t)$, acquired using the ADC. To reduce the noise of our signal, we average each 50 point interval of acquired data into a single point, leading to an effective data sampling rate of 10 MS/s (effective sampling time of 100 ns). Following this averaging process, the data is digitally demodulated, as well as low-pass filtered (--3 dB bandwidth of $\sim$1.2 MHz, time constant $\tau_0 \approx$ 0.8 $\mu$s) around the frequency of interest, $\omega$, via convolution with a Blackman window, $H(t)$. Mathematically, this is interpreted as the ``band-passed'' Fourier transform 
\begin{equation}
\mathcal{V}(\omega,t) = \int_{-\infty}^{\infty} v(t-t') e^{-i \omega (t-t')} H(t') d t',
\label{FFTv}
\end{equation}
performed at each time step, $t$, of the ADC signal. Note that the $\sim$1.2 MHz bandwidth of the filter function is much larger than the linewidth of any of the studied mechanical modes, ensuring that the entire area of each considered resonance peak will be encapsulated. Furthermore, while the data is taken with an effective time step of 100 ns, this filter will smooth over any features that evolve faster than its 0.8 $\mu$s time constant. From the Fourier transform in Eq.~\eqref{FFTv}, we can determine the time-resolved, band-passed power spectral density of $v(t)$ as
\begin{equation}
S_v( \omega,t) = \frac{\left| \mathcal{V}(\omega,t) \right|^2}{\tau_0}.
\label{PSDv}
\end{equation}
If we choose our demodulation frequency to be equal to one of our mechanical resonances ({\it i.e.}, $\omega = \omega_{\rm m}$), this spectral density will provide a direct measure of the mode's phonon occupancy as $\braket{n}(t) \approx k_{\rm B} T_{\rm m} (t) / \hbar \omega_{\rm m} \propto S_x(\omega_{\rm m},t) \propto S_v(\omega_{\rm m},t)$, where $S_x(\omega,t)$ is the band-passed power spectral density of the mechanical mode's displacement, $x(t)$, and $T_{\rm m}(t)$ is the time-dependent mechanical mode temperature (see Appendix \ref{mechT} below).

\subsection{Optomechanical Detection Efficiency}
\label{optoeff}

To determine the overall efficiency of our optomechanical detection, we analyze the losses at each junction of our optical circuit. While coupling to the device, light from the tapered optical fiber is scattered off the substrate, as well as lost as photons travel through the fiber and out of the fridge, with corresponding transmission efficiencies of $\eta_{\rm s}$ = 62.6\% and $\eta_{\rm f} =$ 72.0\%, respectively. Further losses in the fiber at room temperature result in a fraction $\eta_{\rm RT} =$ 81.6\% of the light that exits the fridge reaching the BPD. Including the quantum efficiency of the BPD itself, $\eta_{\rm BPD}$ = 78.1\%, the total optomechanical detection efficiency of the system ({\it i.e.}, the fraction of photons coupled out of the device that are converted into measured photoelectrons) is given by $\eta = \eta_{\rm s} \eta_{\rm f} \eta_{\rm RT} \eta_{\rm BPD} =$ 28.7\%.

\subsection{Thermometry and Temperature Control}
\label{thermo}

To measure the temperature of the base plate of the dilution refrigerator, two complementary thermometers are used. The counts of gamma ray emission from a $^{60}$Co nuclear orientation (NO) thermometer over a 570 s time window, referenced to a high temperature count rate at 4.2 K, provided accurate temperature readings below 50 mK, while the resistance curve of a RuO thermometer is used for $T \ge$ 50 mK. Uncertainty in the temperature readings of the NO thermometer are obtained as the standard deviation in the spread of reported temperatures over the course of a measurement, while the RuO error is taken as the uncertainty in the accuracy of the sensor as specified by the supplier.

In order to heat the dilution refrigerator above its base temperature of 10 mK, current is applied to a resistive heater mounted on the mixing chamber plate, with temperature stability for the duration of a given measurement ensured by a PID-controlled feedback loop referenced to the RuO thermometer. In the range of 10 mK to 800 mK, the cooling power is provided by operating the dilution unit, while for temperatures up to 4.2 K, fridge circulation is ceased and cooling is supplied by the 1K pot. Finally, above 4.2 K, the 1K pot is stopped, such that connection to the liquid helium bath surrounding the fridge is the source of cooling for the base plate.

\section{Optomechanical Device Properties}
\label{optoprop}

The studied optomechanical device, as seen in Fig.~\ref{fig1}(d), is comprised of a suspended half-ring mechanical resonator (width $w =$ 200 nm, thickness $d =$ 250 nm) side-coupled to a 10 $\mu$m diameter optical microdisk cavity, with a vacuum gap of 75 nm between the optical and mechanical elements. The microdisk cavity supports a number of optical modes, each with a resonant frequency $\omega_p$, such that the total electric field in the disk can be expressed as $\vec{\mathscr{E}}(\vec{r},t) = \sum \mathcal{E}_p(t)\vec{\mathscr{E}}_p(\vec{r})$ \cite{eichenfield_2009}. Likewise, the motion of the half-ring resonator can be broken down into a set of mechanical modes at frequency $\omega_q$, such that its displacement from equilibrium can be described by $\vec{u}(\vec{r},t) = \sum x_q(t)\vec{u}_q(\vec{r})$ \cite{eichenfield_2009,hauer_2013}. Here, we have separated both the electric field and displacement profiles into their time-dependent amplitudes, $\mathcal{E}_p(t)$ and $x_q(t)$, with their spatially-varying modeshapes, $\vec{\mathscr{E}}_p(\vec{r})$ and $\vec{u}_q(\vec{r})$, normalized such that ${\rm max}|\sqrt{\epsilon(\vec{r})} \vec{\mathscr{E}}_p(\vec{r})| = {\rm max}|\vec{u}_q(\vec{r})| = 1$ for all $p$ and $q$. In this way, we can characterize the spatial extent of each optical mode through the effective optical mode volume, $V_{\rm opt} = \int \epsilon(\vec{r}) |\vec{\mathcal{\mathscr{E}}}_p(\vec{r})|^2 dV$ \cite{eichenfield_2009,vahala_2004}, as well as the extended nature of the mechanical resonator via an effective mass, $m_q = \int \rho(\vec{r}) |\vec{u}_q(\vec{r})|^2 dV$, for each mechanical mode \cite{eichenfield_2009,hauer_2013}. Both integrals are performed over the volume of the entire optomechanical system, where $\epsilon(\vec{r})$ and $\rho(\vec{r})$ are the dielectric profile and mass density of the device. 

In the experiment, we excite a single optical whispering gallery mode of the resonator at frequency $\omega_p = \omega_{\rm c}$, allowing us to observe its interaction with the mechanical degrees of freedom of the system. Here, a dispersive optomechanical coupling is realized by the fact that the motion of the half-ring resonator perturbs the boundaries of the microdisk cavity, shifting its resonant frequency. To first order, these shifts can be expressed for a given mechanical mode as $\omega_{\rm c}(x_q(t)) \approx \omega_{\rm c} + G x_q(t)$, where $G = d \omega_{\rm c} / dx_q$ is the optomechanical coupling coefficient. Using a perturbative approach \cite{johnson_2002,eichenfield_2009}, an expression for $G$ can be found as
\begin{equation}
G = \frac{\omega_{\rm c}}{2V_{\rm opt}}\displaystyle \int \vec{u}_q(\vec{r}) \cdot \hat{n}(\vec{r}) \left[\Delta \epsilon | \vec{\mathscr{E}}_{p,\parallel}(\vec{r})|^2 - \Delta \epsilon^{-1} \epsilon^2(\vec{r})| \vec{\mathscr{E}}_{p,\perp}(\vec{r})|^2 \right] dA,
\label{Geq}
\end{equation}
where the integration is performed over the surface of the mechanical resonator. Here, $\vec{\mathscr{E}}_{p,\parallel}(\vec{r})$ and $\vec{\mathscr{E}}_{p,\perp}(\vec{r})$ are the components of $\vec{\mathscr{E}}_p(\vec{r})$ parallel and perpendicular to the surface of the mechanical resonator, as defined by its spatially-varying unit normal vector $\hat{n}(\vec{r})$, with $\Delta \epsilon = \epsilon_{\rm d} - \epsilon_{\rm s}$ and $\Delta \epsilon^{-1} = \epsilon_{\rm d}^{-1} - \epsilon_{\rm s}^{-1}$ defined in terms of the permittivities of the device's material, $\epsilon_{\rm d}$, and surrounding medium, $\epsilon_{\rm s}$. We also introduce the single photon, single phonon optomechanical coupling rate, $g_0 = G x_{\rm zpf}$, which corresponds to the shift in the optical cavity's resonance frequency due to the fluctuation amplitude of the resonator's zero-point motion, $x_{\rm zpf} = \sqrt{\hbar / 2 m_q \omega_q}$ ({\it i.e.}, the root-mean-square value of $x_q(t)$ when the mechanical resonator is in its ground state).

Due to optical heating of the mechanical mode, as well as an anomalous optomechanical damping effect that will be the subject of future studies, it is difficult to obtain an experimentally-determined value of $G$. However, by performing FEM simulations of the optical and mechanical modeshapes of the device, we can calculate a value for $G$ according to Eq.~\eqref{Geq}. Furthermore, we use this simulated mechanical modeshape to determine $m_q$, from which we also find $x_{\rm zpf}$, and subsequently, $g_0$. These values for each mechanical mode are found in Table \ref{OMtab}. We note that due to the symmetry of the displacement with respect to the optical field, we simulate $G \approx 0$ for the two lower frequency mechanical modes, even though this symmetry is broken in the experiment, such that significant optomechanical coupling exists.

\begin{table}[h!]
\begin{tabular}{ ccccc }
\hline
$\omega_{\rm m}/2 \pi$ & $m_q$ & $x_{\rm zpf}$ & $G / 2 \pi$ & $g_0 / 2 \pi$ \\  
 (MHz) & (fg) & (fm) & (GHz/nm) & (kHz) \\ \hline
\hline
3.53 & 610 & 62.4 & -- & -- \\

6.28 & 836 & 40.0 & -- & -- \\

15.44 & 743 & 27.0 & 2.34 & 63.4 \\

18.31 & 772 & 24.4 & 5.13 & 125 \\

\hline
\end{tabular}
\caption{Summary of the optomechanical properties for each mechanical mode. The effective mass, $m_q$, and optomechanical coupling coefficient, $G$, are determined using FEM simulations for the electric field and displacement profiles of the optomechanical device. From these values, the zero-point fluctuation amplitude, $x_{\rm zpf}$, and the single-photon optomechanical coupling rate, $g_0$, are also calculated. Note that due to the symmetry of the simulated system, the values of $G$ given here represent a lower bound for the considered geometry and are only nonzero for the two higher frequency mechanical modes.}
\label{OMtab}
\end{table}

\section{Mechanical Mode Temperature}
\label{mechT}

\subsection{Calibration by Varying Fridge Temperature}
\label{calT}

The measured signal in our experiment is a fluctuating voltage, $v(t)$, at the output of a balanced photodetector, which encodes the mechanical motion of our device. Transforming this signal into the frequency domain, we obtain its single-sided, band-passed spectral density function (see Appendix \ref{expdet}), which will in general be given by \cite{hauer_2013}
\begin{equation}
S_v(\omega,t) = \displaystyle \sum_q \alpha_q(\omega,t) S_x^q(\omega,t) + S_v^{\rm imp}(\omega),
\label{Sv}
\end{equation}
where $\alpha_q(\omega,t)$ is the transduction coefficient for the single-sided, band-passed displacement spectral density $S_x^q(\omega,t)$ corresponding to the $q$th mechanical mode of the resonator and $S_v^{\rm imp}(\omega)$ is the frequency-dependent imprecision noise floor of the measurement. If we consider a finite bandwidth, $\Delta \omega$, surrounding the resonance frequency of a single mechanical mode, the sum in Eq.~\eqref{Sv} collapses and we can approximate the transduction coefficient and noise floor as constant over this frequency range. Furthermore, the time-dependence in $\alpha_q(\omega,t)$ is due to the ring-up of the optical cavity, which occurs on a timescale of $1 / \kappa \approx 1$ ns, much faster than any other component in our detection system. We can therefore treat $\alpha_q(\omega,t)$ as a step function in time, such that it takes on a constant value once the laser reaches the optical cavity. We then have $\displaystyle \sum_q \alpha_q(\omega,t) S_x^q(\omega,t) \approx\alpha S_x(\omega,t)$ and $S_v^{\rm imp}(\omega) \approx S_v^{\rm imp}$ resulting in
\begin{equation}
S_v(\omega,t) \approx \alpha S_x(\omega,t) + S_v^{\rm imp},
\label{Svsing}
\end{equation}
where $S_x(\omega,t)$ is the displacement spectral density of the mechanical mode of  interest, with resonant frequency $\omega_q = \omega_{\rm m}$. 

In general, the coefficient $\alpha$ is a combination of a number of experimental parameters such that it is difficult to determine {\it a priori}. We therefore look for a simple way to relate the spectral density of our measurement to the temperature of the mode in question. This is done by relating the spectrum of the mechanical mode to its time-dependent temperature, $T_{\rm m}(t)$, using the expression \cite{hauer_2015}
\begin{equation}
\int S_x(\omega,t) d \omega \approx \frac{4 \pi  x_{\rm zpf}^2 k_{\rm B}}{\hbar \omega_{\rm m}} T_{\rm m}(t),
\label{SxtoT}
\end{equation}
where the integration is performed over the bandwidth $\Delta \omega$ centered on $\omega_{\rm m}$ and we have assumed the experimentally-relevant high-temperature regime ({\it i.e.}, $k_{\rm B} T_{\rm m}(t) \gg  \hbar \omega_{\rm m}$ for all $t$). Combining Eq.~\eqref{SxtoT} with Eq.~\eqref{Svsing}, we find that
\begin{equation}
\int S_v(\omega,t) d \omega = \chi T_{\rm m}(t) + \zeta,
\label{SvtoT}
\end{equation}
indicating that the area under the curve of the measured voltage spectral density is linearly related to the mechanical mode temperature, with proportionality $\chi = 4 \pi x_{\rm zpf}^2 k_{\rm B} \alpha / \hbar \omega_{\rm m}$ and a constant offset $\zeta = S_v^{\rm imp} \Delta \omega$ set by the noise floor and bandwidth of the measurement. Fitting Eq.~\eqref{SvtoT} to the area under the voltage spectral density at the beginning of the pulse, when the mechanical mode is thermalized to the fridge temperature, $T_{\rm f}$, at $t=t_0$ allows us to infer the initial mode temperature as $T_{\rm m}(t_0) = T_{\rm f}$ and extract values for $\chi$ and $\zeta$. Provided the conditions stay the same throughout the experiment, we can use these parameters to determine the mode temperatures at later times in the pulse as the device rapidly heats due to interaction with the photon-induced bath. An example of this type of calibration is seen in Fig.~\ref{fig2}.

\subsection{Heating Effects}
\label{heateff}

In the previous section, we assumed that the device is initially thermalized to the mixing chamber of the fridge so that their temperatures are identical. However, due to diminishing thermal conductivities at low temperatures, this may not be the case. To investigate this, we use a simple model to estimate the chip temperature (in the vicinity of the device) for varying average input powers to the chip.

\begin{figure}[t!]
\centerline{\includegraphics[width=3.0in]{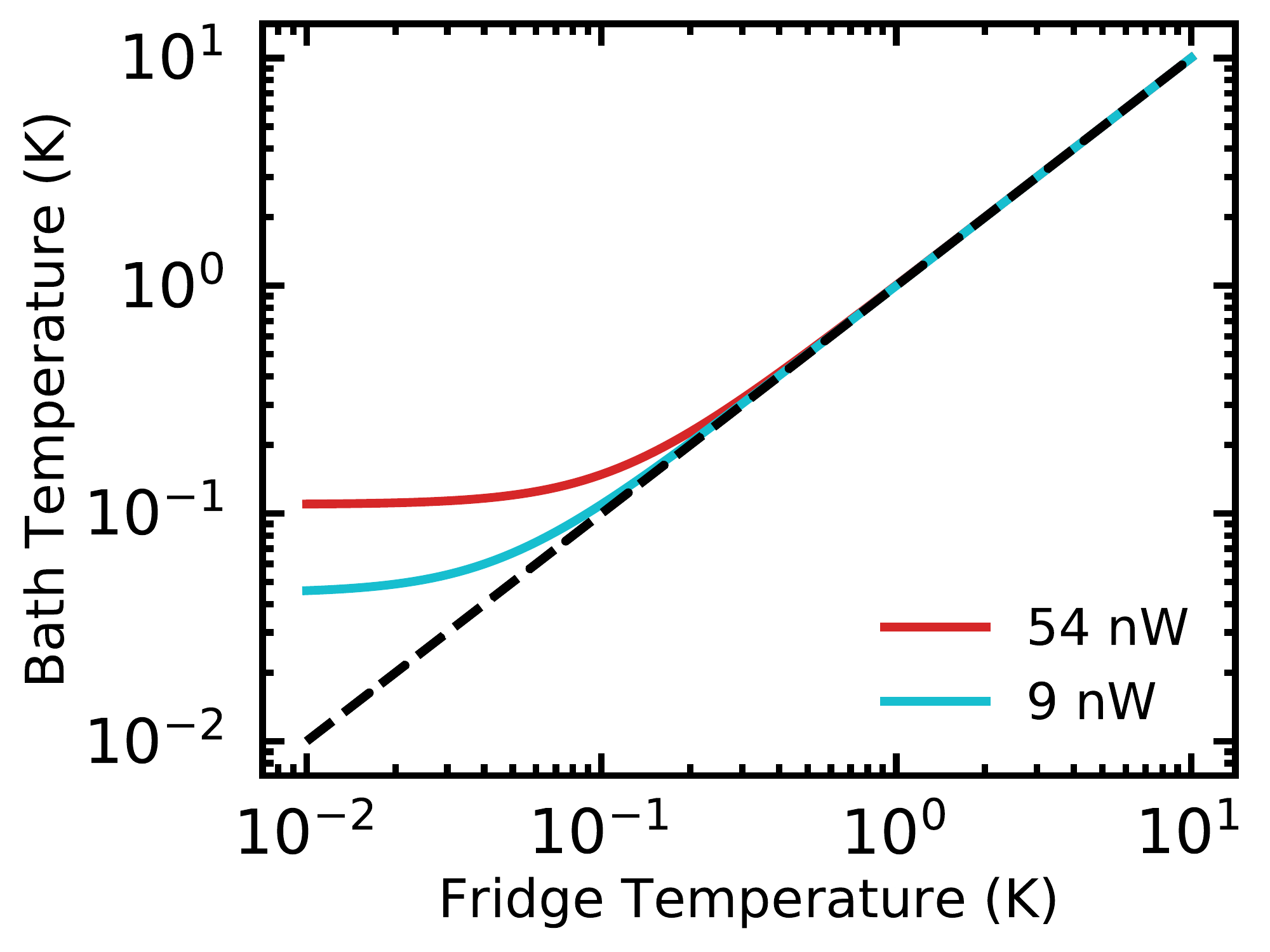}}
\caption{{\label{fig6}} Plot of device's bath temperature, $T_{\rm b}$, vs fridge/chip-holder temperature, $T_{\rm f}$, according to Eq.~\eqref{devtemp} for the largest (54 nW) and smallest (9 nW) estimated average heat loads, $\dot{Q}_{\rm d}$, applied to the device during pulsed measurements. The black dashed line is that of a perfectly thermalized device, {\it i.e.}, $T_{\rm b} = T_{\rm f}$.}
\end{figure}

We begin by assuming that the chip holder is well thermalized to the fridge [via a large copper braid -- see Fig.~\ref{fig1}(b)] such that it's temperature is equal to that of the base plate. Furthermore, we assume that the cooling power of the mixing chamber is large enough that this temperature remains constant over the course of the measurement (as this is what we observe during the experiment). The bath temperature of our device, $T_{\rm b}$, is then limited by the thermal conductivity of the 210 nm thick gold layer applied to the top of our silicon device layer to improve the thermal conductivity of our chip at low temperatures. We note that $T_{\rm b}$ is the temperature of the chip in the vicinity of our device ({\it i.e.}, the temperature of it's thermal bath), as opposed to the mechanical mode temperature, $T_{\rm m}$, introduced in the previous section. Assuming the thermal conductivity of the gold layer varies linearly at low temperatures as $k_{\rm g} = k_0 T$, where $k_0 \approx$ 30 W/m$\cdot$K$^2$ \cite{white_1953}, the device's bath temperature will be given by
\begin{equation}
T_{\rm b} = \sqrt{\frac{2 L_{\rm g} \dot{Q}_{\rm d}}{A_{\rm g} k_0} + T_{\rm f}^2}.
\label{devtemp}
\end{equation}
Here, $\dot{Q}_{\rm d}$ is the heat load applied to the device, with $L_{\rm g}$ and $A_{\rm g}$ being the length and cross-sectional area of the thermalizing gold layer.
For a double pulse measurement at a power of $P_{\rm f} =$ 10 $\mu$W input to the fridge, as was done during the measurement, the fraction of power absorbed at the chip can be approximated as $P_{\rm abs} \approx \eta_f (1 - \eta_s) P_{\rm f} =$ 2.7 $\mu$W. Including the duty cycle of the measurement, which for pulse delay times much less than the 200 ms wait time per measurement can be approximated as 2 $\times$ 2 ms$/$200 ms $\approx$ 0.02, we get an average power applied to the device during measurement of $\dot{Q}_{\rm d} \approx$ 54 nW. Conversely, for our longest wait time of 1 s, the duty cycle decreases to 2 $\times$ 2 ms$/$1200 ms $\approx$ 0.003, leading a lower average measurement power of $\dot{Q}_{\rm d} \approx$ 9 nW. 

Fig.~\ref{fig6} shows the values of $T_{\rm b}$ according to Eq.~\eqref{devtemp} versus fridge/chip holder temperature, $T_{\rm f}$, for each of these two heat loads input to the device, where we have taken $L_{\rm g} =$ 3.5 mm and $A_g =$ 5 mm $\times$ 210 nm = 1.05$\times$10$^{-9}$ m$^2$ according to the experiment. As can be seen, for these applied heat loads, the device is no longer thermalized to the fridge at temperatures $T \lesssim$ 100 mK. We note that this treatment neglects a number of effects, such as a Kapitza boundary resistance \cite{pobell_2007} between each interface of the apparatus, as well as the relevant time scales associated with the measurements and heating/cooling processes. Nonetheless, this simple model provides evidence that the low temperature plateau in the damping rates seen in Fig.~\ref{fig4} may be due to measurement-induced heating of the device's thermal environment.

\section{Strain Energy Distribution in Mechanical Resonators}
\label{strain}

\subsection{Effective Strain Volume}
\label{Vstr}

As mentioned in the main text, it is the strain induced by the motion of the mechanical resonator that couples to defects in the device's material. It is therefore important to understand the spatially-varying strain profiles of each mechanical mode. To do this, we begin with the total (time-dependent) strain energy density of the resonator, which includes all mechanical modes and is given by \cite{achenbach_1973, sadd_2005}
\begin{equation}
U(\vec{r},t) = \frac{1}{2} C_{abcd} \varepsilon_{ab}(\vec{r},t) \varepsilon_{cd}(\vec{r},t),
\label{strenergydens}
\end{equation}
such that the total (time-dependent) elastic potential energy of the system is
\begin{equation}
E_{\rm p}(t) = \int U(\vec{r}) dV,
\label{Estore}
\end{equation}
where the integral is performed over the entire volume of the solid in question and we have used standard Einstein summation notation ({\it i.e.}, sum over repeated indices). Here, $C_{abcd}$ are the components of the elasticity tensor $\tensor{C}$ of the material, while $\varepsilon_{ab}(\vec{r},t)$ are the compoents of the strain tensor $\tensor{\varepsilon}(\vec{r},t)$ induced by the mechanical motion, which can be expressed as
\begin{equation}
\tensor{\varepsilon}(\vec{r},t) = \sum_q \xi_q(t) \tensor{\varepsilon}_q(\vec{r}).
\label{straindef}
\end{equation}
As we did with the mechanical displacement in Appendix \ref{optoprop}, we have broken this strain tensor into its time-varying amplitude, 
$\xi_q(t) = \sqrt{m_q} x_q(t)$, and spatially-varying strain profile tensor, $\tensor{\varepsilon}_q(\vec{r})$, the components of which are given by $\varepsilon_{q,ab}(\vec{r}) = \frac{1}{2 \sqrt{m_q}} (\partial u_{q,a} / \partial \tilde{x}_b + \partial u_{q,b} / \partial \tilde{x}_a )$, with $u_{q,i}(\vec{r},t)$ and $\tilde{x}_i$ being the $i$th components of the mechanical displacement and coordinate vectors.

For systems that exhibit cubic symmetry, such as the diamond lattice of silicon, the elasticity tensor has only three independent, nonzero components, namely $C_{xxxx} = C_{yyyy} = C_{zzzz} = C_{11} =$ 165.6 GPa, $C_{xxyy} = C_{xxzz} = C_{yyzz} = C_{yyxx} = C_{zzxx} = C_{zzyy} = C_{12} =$ 63.9 GPa and $C_{xyxy} = C_{yxyx} = C_{xyyx} = C_{yxxy} = C_{xzxz} = C_{zxzx} = C_{xzzx} = C_{zxxz} = C_{yzyz} = C_{zyzy} = C_{yzzy} = C_{zyyz} = C_{44} =$ 79.5 GPa, with all other components being zero \cite{mason_1958}. Using these symmetries of the elasticity tensor, the elastic strain energy density for a given mechanical mode, time-averaged over its mechanical oscillation period $\tau_0$, can be expressed as
\begin{equation}
\begin{split}
&U_q(\vec{r}) = \frac{1}{\tau_0} \int_0^{\tau_0} \frac{1}{2} \Big[ C_{11} \left(\varepsilon^2_{q,xx}(\vec{r}) + \varepsilon^2_{q,yy}(\vec{r}) + \varepsilon^2_{q,zz}(\vec{r})\right) \\ &+ 4 C_{44} \left(\varepsilon^2_{q,xy}(\vec{r}) + \varepsilon^2_{q,xz}(\vec{r}) + \varepsilon^2_{q,yz}(\vec{r}) \right) \\ 
&+ 2 C_{12} \big(\varepsilon_{q,xx}(\vec{r})\varepsilon_{q,yy}(\vec{r}) + \varepsilon_{q,xx}(\vec{r})\varepsilon_{q,zz}(\vec{r}) + \varepsilon_{q,yy}(\vec{r})\varepsilon_{q,zz}(\vec{r}) \big) \Big] dt.
\label{strenergydenssimp}
\end{split}
\end{equation}

As seen in Fig.~\ref{fig4}, the majority of the strain energy density is localized to certain portions of the resonator, such that the mechanical motion will only couple to a specific subset of defects that occupy these regions of high strain. To characterize the extent to which each mechanical mode of the resonator probes these defects, we define an effective strain volume \cite{ramos_2013}
\begin{equation}
V_{\rm str} = \int \frac{U_q(\vec{r})}{{\rm max}[U_q(\vec{r})]} dV,
\label{Veff}
\end{equation}
analogous to the effective mode volumes for optical cavities in cavity quantum electrodynamics (see Appendix \ref{optoprop}), where ${\rm max}[U_q(\vec{r})]$ is the maximum value of the strain energy density for the $q$th mechanical mode. The effective strain volumes for each of the four mechanical modes considered in this work are calculated using FEM simulations of their strain energy density profiles (see Fig.~\ref{fig4}) with the smallest mesh allowable (set by computing constraints) and are displayed in Table \ref{TLStab} of Section \ref{quantagree}.

We are further interested in determining the fraction of the strain energy that is localized to the surface of the resonator. To do this, we use FEM simulations to compare the strain energy located within the first 5, 10, 20 and 40 nm of the resonator's surface (note that this corresponds to roughly the first 10, 20, 40, and 80 monolayers of silicon, as its lattice constant is 5.4 nm) to the strain energy of the entire structure for each mechanical mode. The results of these calculations are given in Table \ref{surffractab}. As one can see, over half of the strain energy density is localized to the small volume surrounding the first 20 nm of the resonator (corresponding to roughly 37\% of the resonator's geometric volume) for each mechanical mode, with nearly all of the strain energy being located within 40 nm of the surface, furthering our hypothesis that the mechanical dissipation observed in this work is caused by coupling to surface defects. We also note that there is slightly less strain at the surface for the two lower effective strain volume modes (3.53 MHz, 18.31 MHz) than there is for the two high effective strain volume modes (6.28 MHz, 15.44 MHz), an effect that may contribute in part to the higher defect density probed by these latter two modes.

\begin{table}[h!]
\begin{tabular}{ ccccc }
\hline
$\omega_{\rm m}/2 \pi$ & 5 nm & 10 nm & 20 nm & 40 nm \\ \hline
\hline
3.53 MHz & 0.17 & 0.31 & 0.55 & 0.83 \\

6.28 MHz & 0.21 & 0.38 & 0.64 & 0.91 \\

15.44 MHz & 0.21 & 0.39 & 0.65 & 0.92 \\

18.31 MHz & 0.19 & 0.35 & 0.60 & 0.87 \\

\hline
\end{tabular}
\caption{Fraction of the strain energy localized to within 5, 10, 20 and 40 nm of the resonator's surface for each of the four studied mechanical modes.}
\label{surffractab}
\end{table}

\subsection{Strain Energy Fraction}
\label{strfrac}

\begin{figure}[h!]
\centerline{\includegraphics[width=3in]{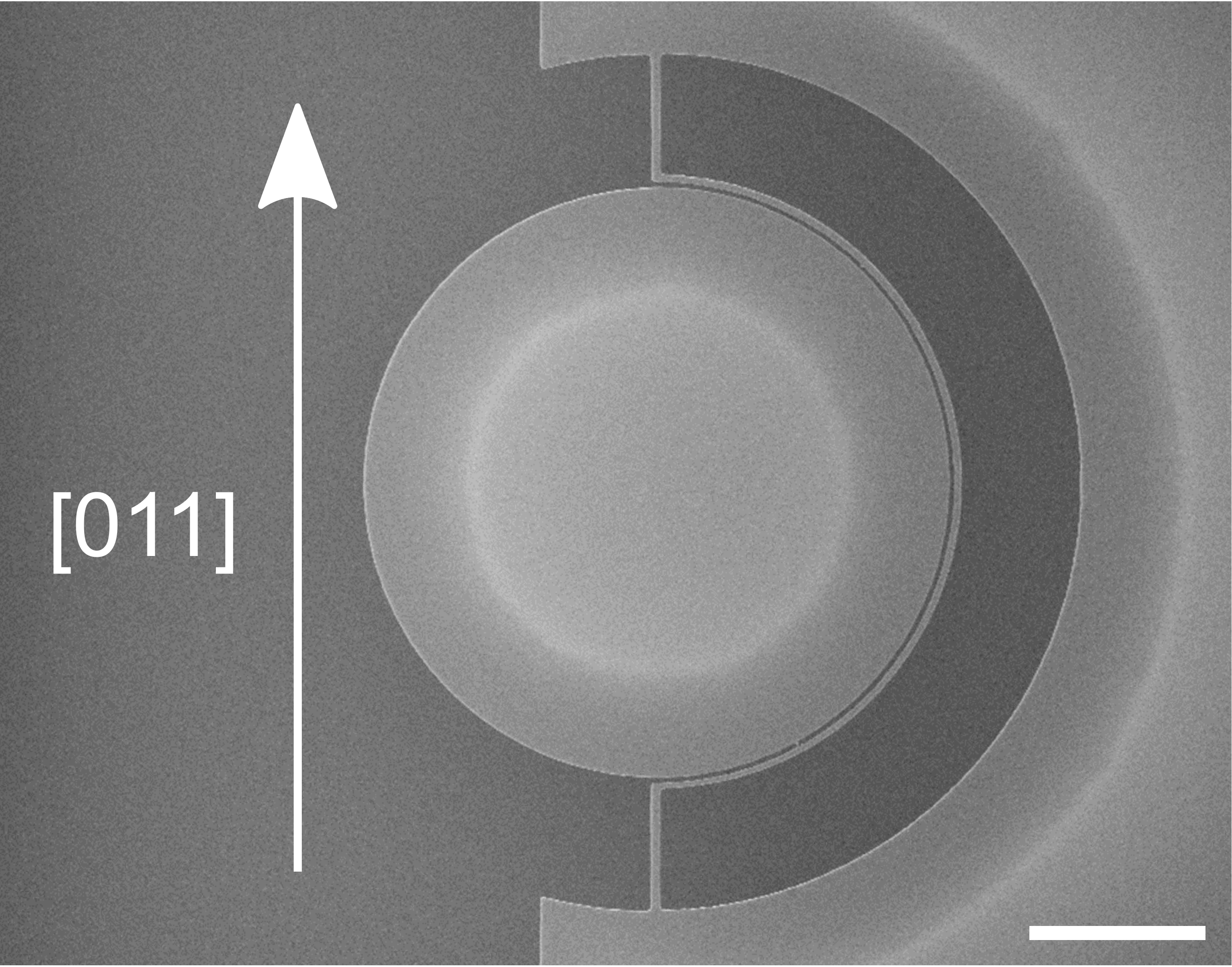}}
\caption{{\label{fig7}} SEM image of the studied device with the orientation of the crystal axis in the silicon device layer specified. Scale bar is 3 $\mu$m.}
\end{figure}

It is also useful to break the total energy of each mechanical mode into fractions corresponding to each of the solid's phononic mode polarizations ({\it i.e.}, to determine to what extent the mode is longitudinal or transverse). To do this, we begin with the total mechanical energy of the resonator, including both kinetic and elastic potential energy, which is given by \cite{behunin_2016}
\begin{equation}
E_{\rm m} = \frac{1}{2} \int \left( \rho(\vec{r}) |\dot{\vec{u}}(\vec{r},t)|^2 + C_{abcd} \varepsilon_{ab}(\vec{r},t) \varepsilon_{cd}(\vec{r},t) \right) dV.
\label{Emech}
\end{equation}
Using the orthonormality of the normal mode representation of the elastic field, we can express this energy as \cite{behunin_2016}
\begin{equation}
\begin{split}
&E_{\rm m} = \sum_q E_q \Bigg\{ \frac{1}{\omega_q^2} \int \rho(\vec{r}) \Bigg[ c_l^2 \sum_{a,b} \varepsilon_{q,aa}(\vec{r})\varepsilon_{q,bb}(\vec{r}) \\ 
&+ c_{t_1}^2 \left( \sum_a |\varepsilon_{q,aa}(\vec{r})|^2 - \sum_{a \ne b} \varepsilon_{q,aa}(\vec{r})\varepsilon_{q,bb}(\vec{r}) \right)  \\
&+ c_{t_2}^2 \left( 2 \sum_{a \ne b} \varepsilon_{q,ab}(\vec{r})\varepsilon_{q,ab}(\vec{r}) - \sum_{a,b} \varepsilon_{q,aa}(\vec{r})\varepsilon_{q,bb}(\vec{r}) \right) \Bigg] \Bigg\} dV,
\label{Emechq}
\end{split}
\end{equation}
where $E_q = \hbar \omega_q ( n_q + 1/2)$ is the energy of the $q$th mechanical mode with phonon occupancy of $n_q$ and
\begin{equation}
\begin{split}
c_l &= \sqrt{\frac{C_{11}+C_{12}+2 C_{44}}{2 \rho}}, \\
c_{t_1} &= \sqrt{\frac{C_{11}-C_{12}}{2 \rho}}, \\
c_{t_2} &= \sqrt{\frac{C_{44}}{\rho}},
\label{soundspeeds}
\end{split}
\end{equation}
are the speeds of sound associated with the longitudinal elastic wave polarized in the $[110]$ direction, as well as the two transverse waves polarized in the $[001]$ and $[1\bar{1}0]$ directions, respectively \cite{mason_1958}. Using $\rho =$ 2330 kg/m$^3$ as the uniform density of silicon, these speeds of sound are found to be $c_l =$ 9148 m/s, $c_{t_1} =$ 4679 m/s and $c_{t_2} =$ 5857 m/s. Upon inspection of Eq.~\eqref{Emechq}, we see that we can define a fraction of the $q$th mode's mechanical energy associated with each of these polarizations as
\begin{equation}
\begin{split}
e_{ql} &= \frac{c_l^2}{\omega_q^2} \int \rho(\vec{r}) \sum_{a,b} \varepsilon_{q,aa}(\vec{r})\varepsilon_{q,bb}(\vec{r}) dV, \\
e_{qt_1} &= \frac{c_{t_1}^2}{\omega_q^2} \int \rho(\vec{r}) \left( \sum_a |\varepsilon_{q,aa}(\vec{r})|^2 - \sum_{a \ne b} \varepsilon_{q,aa}(\vec{r})\varepsilon_{q,bb}(\vec{r}) \right) dV, \\
e_{qt_2} &= \frac{c_{t_2}^2}{\omega_q^2} \int \rho(\vec{r}) \left( 2 \sum_{a \ne b} \varepsilon_{q,ab}(\vec{r})\varepsilon_{q,ab}(\vec{r}) - \sum_{a,b} \varepsilon_{q,aa}(\vec{r})\varepsilon_{q,bb}(\vec{r}) \right) dV.
\label{polfracs}
\end{split}
\end{equation}
We note that since $\sum \varepsilon_{q,aa}(\vec{r})\varepsilon_{q,bb}(\vec{r}) = |{\rm Tr} \{ \tensor{\varepsilon}_q(\vec{r}) \} |^2$, which is an invariant of the strain tensor, $e_{ql}$ remains constant with respect to changes of coordinates ({\it i.e.}, crystal orientations) for a given mechanical mode, however, this is not the case for $e_{qt_1}$ and $e_{qt_2}$. We determine these fractions for the four mechanical modes studied in this work by performing FEM simulations of the strain profiles for each, with the crystal orientation chosen to match that of the device used in the experiment (see Fig.~\ref{fig7}). The results of these calculations can be seen in Table \ref{fractab}.

\begin{table}[h!]
\begin{tabular}{ cccc }
\hline
$\omega_{\rm m}/2 \pi$ & $e_{ql}$ & $e_{qt_1}$ & $e_{qt_2}$ \\ \hline
\hline
3.53 MHz & ~~~~0.15 & ~~~~0.42 & ~~~~0.43 \\

6.28 MHz & ~~~~0.34 & ~~~~0.54 & ~~~~0.12 \\

15.44 MHz & ~~~~0.34 & ~~~~0.52 & ~~~~0.14  \\

18.31 MHz & ~~~~0.39 & ~~~~0.32 & ~~~~0.29  \\

\hline
\end{tabular}
\caption{Mechanical energy fractions for each of the four modes studied in this work. The values are determined using FEM simulations of the strain profile for each mode, with the orientation of the silicon crystal axes chosen to match the device used in the experiment (see Fig.~\ref{fig7}).}
\label{fractab}
\end{table}

\section{Standard Tunneling Model for Acoustic Damping in One-Dimensional Systems}
\label{STMmodel}

In the early 1970's, it was discovered by Zeller and Pohl \cite{zeller_1971} that the cryogenic thermal properties of a number of glassy solids deviated significantly from what was expected according to the Debye model. To account for this anomalous behaviour, Anderson {\it et al.} \cite{anderson_1972} and Phillips \cite{phillips_1972} simultaneously developed what is now known as the {\it standard tunneling model}, whereby phonons in the solid exchange energy with the medium by driving configurational changes of intrinsic defect states. Further extensions to this model were made by J\"ackle {\it et al.}~\cite{jackle_1972,jackle_1976}, who used it to correctly describe the anomalous acoustic absorption observed in fused quartz \cite{heinicke_1971}. 

While early incarnations of the standard tunneling model were used with great success to describe the cryogenic properties of bulk amorphous solids, modifications to this model are necessary in order to account for the behaviour of defect-phonon coupling in reduced dimensionality systems fabricated from crystalline solids \cite{behunin_2016}. Here, we introduce the standard tunneling model in the original form used to model defects in amorphous solids and extend it to describe the mechanical dissipation in crystalline nanomechanical resonators.

\subsection{Double-Well Potential Model for Tunneling Systems}
\label{doublewell}

\begin{figure}[h!]
\centerline{\includegraphics[width=\columnwidth]{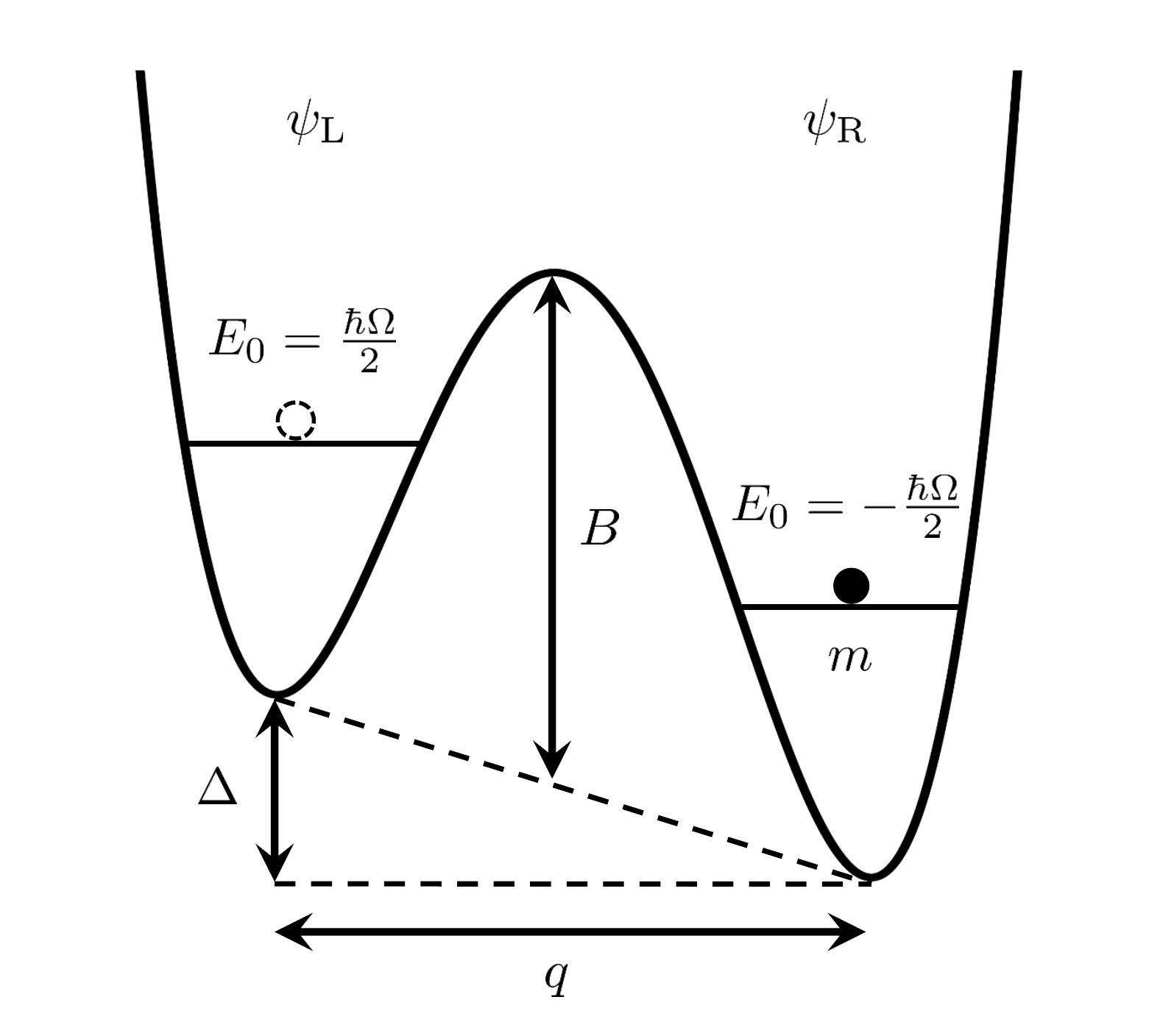}}
\caption{{\label{fig8}} Schematic of a particle of mass, $m$, tunneling between the two ground states of a double-well potential, separated by a barrier of height, $B$, and a generalized configurational coordinate, $q$, with an asymmetry, $\Delta$, between their two minima. The particle can tunnel through the barrier, allowing it to be localized in the ground state of either the left or right well (each with ground state energy $E_0 = \hbar \Omega / 2$), as described by the wavefunctions $\psi_{\rm L}$ and $\psi_{\rm R}$, respectively.}
\end{figure}

In the standard tunneling model, the configurational states of the defects in the solid are modeled as a particle of mass $m$ confined to an asymmetric double-well potential \cite{anderson_1972,phillips_1972}, as seen in Fig.~\ref{fig8}. We assume this potential to be comprised of two identical harmonic wells, each with a ground state energy, $E_0 = \hbar \Omega / 2$, offset by an asymmetry energy, $\Delta$, and separated by a barrier of height, $B$, and the configurational coordinate, $q$. We consider the system to be at low enough temperatures ($k_{\rm B} T \ll \hbar \Omega$) such that only the ground state of each well will be populated with any significant probability. This allows for a two-level system (TLS) description of these two lowest lying configurational states, with wavefunctions $\psi_{\rm L}(\vec{r})$ [$\psi_{\rm R}(\vec{r})$] corresponding to the particle occupying the higher (lower) energy state in the left (right) well. In this set of localized basis states, the Hamiltonian will be given by \cite{phillips_1987, esquinazi_1998}
\begin{equation}
\hat{\mathcal{H}}_{\rm TLS} = \frac{1}{2} \Delta \hat{\sigma}_z - \frac{1}{2} \Delta_0 \hat{\sigma}_x,
\label{Hloc}
\end{equation}
where we have chosen zero energy to be the midway point between the minimum of each well and $\hat{\sigma}_x$ ($\hat{\sigma}_z$) is the $x$ ($z$) Pauli spin matrix. In this Hamiltonian, quantum tunneling between the two states of the TLS is characterized by the {\it tunnel splitting} or {\it tunneling energy}, $\Delta_0$, which can be determined using the Wentzel-Kramers-Brillioun (WKB) approximation to be $\Delta_0 \approx \hbar \Omega e^{-\lambda}$, where $\lambda = \sqrt{2 m B q^2 / \hbar^2}$ is known as the {\it tunneling} or {\it Gamow parameter} and characterizes the penetration of the wavefunctions into the barrier \cite{phillips_1987}.

The Hamiltonian in Eq.~\eqref{Hloc} can be diagonalized by rotating the basis by an angle $\varphi$ defined by $\tan(2 \varphi) = \Delta_0 / \Delta$, resulting in the new Hamiltonian \cite{phillips_1987,esquinazi_1998}
\begin{equation}
\hat{H}_{\rm TLS} = \frac{1}{2} E \hat{\sigma}_z,
\label{Hdiag}
\end{equation}
in the energy eigenstate basis
\begin{equation}
\begin{split}
\psi_+(\vec{r}) & = \psi_{\rm L}(\vec{r}) \cos(\varphi) - \psi_{\rm R}(\vec{r}) \sin (\varphi), \\
\psi_-(\vec{r}) & = \psi_{\rm L}(\vec{r}) \sin(\varphi) + \psi_{\rm R}(\vec{r}) \cos (\varphi).
\label{energyeigstates}
\end{split}
\end{equation}
Here, $E = \sqrt{\Delta^2 + \Delta_0^2}$ is the energy separation between the two states of the TLS, with the wavefunctions $\psi_\pm(\vec{r})$ corresponding to the eigenvalues $\pm E / 2$. If the TLS is in thermal equilibrium with a bath at a temperature $T$, we can use the diagonalized Hamiltonian of Eq.~\eqref{Hdiag} to determine the probability that the TLS is in either of its two states as
\begin{equation}
p^0_\pm = \frac{e^{\mp E / 2 k_{\rm B} T}}{e^{E / 2 k_{\rm B} T}+e^{-E / 2 k_{\rm B} T}} = \frac{1}{e^{ \pm E / k_{\rm B} T}+1},
\label{TLSprob}
\end{equation}
with $p^0_+$ ($p^0_-$) corresponding to the excited (ground) state. From these probabilities, we also define a population inversion probability as
\begin{equation}
s^0 = p^0_+ - p^0_- = -\tanh\left( \frac{E}{2 k_{\rm B}T} \right).
\label{invprob}
\end{equation}

\subsection{Coupling to Phononic Systems}
\label{phoncoup}

Tunneling systems that are embedded in a solid are able to exchange energy with the various excitations of the surrounding medium. Here, we focus on insulating solids, such that the dominant excitation at low temperatures will be quantized vibrations of the lattice, {\it i.e.}, phonons. If the interacting phonon has energy on the order of, or greater than, the TLS separation energy, it can be directly absorbed, promoting a TLS in its ground state to its excited state. However, for the temperatures (10 mK to 10 K) considered in this experiment, TLS at the frequencies relevant for this resonant interaction ($<$ 20 MHz) will be thermally saturated such that absorption or emission of a phonon is equally likely \cite{behunin_2016}. Therefore, this dissipation mechanism does not need to be considered for the MHz frequency mechanical modes studied in this work. Instead, we focus on another TLS-phonon interaction, known as the {\it relaxation interaction} \cite{phillips_1987, esquinazi_1998, enss_2005}, whereby non-resonant phonons generate strains that perturb the local TLS environment, driving the system out of thermal equilibrium by shifting the energy separation between their two levels. This allows the TLS to interact with the lower frequency vibrational modes of the solid, absorbing and emitting phonons until it can relax back to thermal equilibrium. 

To model this relaxation effect, we consider the full Hamiltonian for the interaction between the modes of a mechanical resonator and an ensemble of TLS defects, given by the so-called ``spin-boson'' Hamiltonian \cite{behunin_2016,leggett_1987,seoanez_2008}
\begin{equation}
\begin{split}
\hat{H} &= \sum_q \hbar \omega_q \hat{b}^\dag_q \hat{b}^{\phantom{\dag}}_q + \frac{1}{2} \sum_j E_j \hat{\sigma}_z \\ 
&+ \sum_j \left( \frac{\Delta_{0j}}{E_j}\hat{\sigma}_x + \frac{\Delta_{j}}{E_j}\hat{\sigma}_z \right) \tensor{\gamma_j}:\tensor{\varepsilon} + \hat{H}_{\Gamma}.
\label{Htot}
\end{split}
\end{equation}
In this Hamiltonian, the first two terms correspond to the energies of the resonator's mechanical modes, each with angular frequency $\omega_q$ and annihilation (creation) operator $\hat{b}^{\phantom{\dag}}_q$ ($\hat{b}^\dag_q$), and the TLS ensemble, with a tunneling, asymmetry and separation energy of $\Delta_{0j}$, $\Delta_j$ and $E_j = \sqrt{\Delta_j^2+\Delta_{0j}^2}$ for each TLS. The third term then describes the coupling between the TLS ensemble and the mechanical motion of the resonator, characterized by the dyadic (tensor) product $\tensor{\gamma_j}:\tensor{\varepsilon} = \gamma_{j,ab} \varepsilon_{ab}$ between the deformation potential tensor ({\it i.e.}, the strain-TLS coupling tensor) $\tensor{\gamma_j}$ of the $j$th TLS and the strain tensor $\tensor{\varepsilon}$ induced by the resonator motion \cite{esquinazi_1998,anghel_2007,behunin_2016}. Using Eq.~\eqref{straindef}, along with the fact that the (quantized) displacement amplitude of each mechanical mode can be expressed as $\hat{x}_q = x_{\rm zpf} ( \hat{b}^{\phantom{\dag}}_q + \hat{b}^\dag_q)$, we can write the system Hamiltonian in the more succinct form
\begin{equation}
\begin{split}
\hat{H} &= \sum_q \hbar \omega_q \hat{b}^\dag_q \hat{b}^{\phantom{\dag}}_q + \frac{1}{2} \sum_j E_j \hat{\sigma}_z \\
&+ \sum_j \sum_q \left(\mu_{qj} \hat{\sigma}_x + \nu_{qj}\hat{\sigma}_z \right) (\hat{b}^{\phantom{\dag}}_q + \hat{b}^{\dag}_q) + \hat{H}_{\Gamma},
\label{Htotsuc}
\end{split}
\end{equation}
where we have introduced the TLS-phonon coupling coefficients $\mu_{qj} = \frac{\Delta_{0j}}{E_j} \sqrt{\frac{\hbar}{2 \omega_q}} \tensor{\gamma_j} :\tensor{\varepsilon}_q(\vec{r}_j)$ and $\nu_{qj} = \frac{\Delta_j}{E_j} \sqrt{\frac{\hbar}{2 \omega_q}} \tensor{\gamma_j}:\tensor{\varepsilon}_q(\vec{r}_j)$. We note that for each of these coefficients, the strain is evaluated at the position of the $j$th TLS, denoted by the position vector $\vec{r}_j$. Finally, $\hat{H}_\Gamma$ describes the interaction of the resonator with its environmental bath, which accounts for dissipation mechanisms aside from those due to TLS-phonon interactions, as well as the thermal drive of the mechanical motion.

Coupling between the mechanical modes of the resonator and the TLS ensemble as described by the the Hamiltonian in Eq.~\eqref{Htotsuc} will act to shift the energy separation of each TLS in time according to
\begin{equation}
E_j'(t) = E_{j,+}(t) - E_{j,-}(t) = E_j + \delta E_j(t),
\label{pertE}
\end{equation}
with
\begin{equation}
\delta E_j(t) = 2 \sum_q \nu_{qj} \left( \beta_q (t) + \beta_q^*(t) \right),
\label{delE}
\end{equation}
where we have introduced $\beta_q(t) = \braket{\hat{b}_q(t)}$. This shift in the separation energy will additionally act to perturb the difference in population between the excited and ground state of each TLS away from equilibrium, leading to a time-dependent inversion probability 
\begin{equation}
s_{j}(t) = \braket{\sigma_z} = p_{j,+}(t) - p_{j,-}(t) = s^0_j + \delta s_j(t),
\label{TLSprobinst}
\end{equation}
where $\delta s_j(t)$ is the instantaneous deviation of the inversion probability away from its equilibrium value, $s^0_j$, in the absence of the phonon-induced strain.

In order to determine $\delta s_j(t)$, we must first realize that the perturbed system will strive towards a new, time-dependent equilibrium inversion probability, $\bar{s}_j(t) = \bar{p}_{j,+}(t) - \bar{p}_{j,-}(t)$, which can be found by inputting Eq.~\eqref{pertE} into the expression for $s^0_j$ in Eq.~\eqref{invprob} and expanding to first order to obtain 
\begin{equation}
\begin{split}
\bar{s}_j(t) &= s^0_j + \frac{ds^0_j}{dE_j} \delta E_j(t) \\
&= s^0_j - {\rm sech}^2 \left( \frac{E_j}{2 k_{\rm B} T} \right) \frac{\delta E_j(t)}{2 k_{\rm B} T}.
\label{TLSprobinsteq}
\end{split}
\end{equation}
This ``instantaneous'' equilibrium probability can be interpreted as the inversion probability that the system would reach if the TLS energy separation stayed at $E_j'(t)$ for a sufficiently long time. However, a given TLS cannot immediately achieve this new equilibrium, as it must do so by exchanging energy with the surrounding phonon bath, such that the probabilities of the excited and ground states evolve according to \cite{phillips_1987}
\begin{equation}
\begin{split}
\dot{p}_{j,+} &= -p_{j,+} \upsilon_{j,-} + p_{j,-} \upsilon_{j,+}, \\
\dot{p}_{j,-} &= p_{j,+} \upsilon_{j,-} - p_{j,-} \upsilon_{j,+},
\label{TLSprob+-}
\end{split}
\end{equation}
where $\upsilon_{j,-}$ ($\upsilon_{j,+}$) is the phonon-induced transition rate associated with the excitation (de-exictation) of the TLS. By examining the steady state of Eq.~\eqref{TLSprob+-}, we can see that these transition rates obey the condition of detailed balance, such that $\upsilon_{j,+}/\upsilon_{j,-} = p^0_{j,+}/p^0_{j,-} = \bar{p}_{j,+}(t)/\bar{p}_{j,-}(t) = e^{-E_j/k_{\rm B} T}$ \cite{esquinazi_1998, jackle_1976}. Using this relation, along with the conservation of probability, $p_{j,+}(t) + p_{j,-}(t) = 1$, we find
\begin{equation}
\begin{split}
\dot{s}_j &= -(p_{j,+} - p_{j,-})(\upsilon_{j,+} + \upsilon_{j,-}) + \upsilon_{j,+} - \upsilon_{j,-} \\
&= - \frac{s_j - \bar{s}_j}{\tau_j},
\label{TLSprob+decay}
\end{split}
\end{equation}
where we have introduced the relaxation rate of the TLS populations as
\begin{equation}
\tau^{-1}_j = \upsilon_{j,+} + \upsilon_{j,-} = \upsilon_{j,-} (e^{-E_j/k_{\rm B} T} + 1).
\label{relaxrate}
\end{equation}
This rate can be interpreted as the inverse of the relaxation time, $\tau_j$, required for the inversion probability of a given TLS to relax back to its steady-state value after it has been perturbed away from equilibrium. By inputting Eq.~\eqref{TLSprobinsteq} into Eq.~\eqref{TLSprob+decay}, while using the fact that $\dot{s}_j(t) = \dot{\delta s}_j(t)$, we find
\begin{equation}
\tau_j \dot{\delta s}_j = -\delta s_j - \frac{1}{2 k_{\rm B} T} {\rm sech}^2 \left( \frac{E_j}{2 k_{\rm B} T} \right) \delta E_j(t),
\label{TLSdevdecay}
\end{equation}
which can be Fourier transformed to obtain
\begin{equation}
\delta s_j(\omega) = - \frac{1}{2 k_{\rm B} T}{\rm sech}^2 \left( \frac{E_j}{2 k_{\rm B} T} \right) \frac{\delta E_j(\omega)}{1 - i \omega \tau_j},
\label{TLSdevdecayfreq}
\end{equation}
resulting in the frequency domain solution for the deviation of the inversion probability from equilibrium.

We now look to find an expression for the TLS relaxation rate given by Eq.~\eqref{relaxrate}. This can be done by applying a Fermi's Golden Rule calculation using the interaction Hamiltonian [{\it i.e.}, the third term in Eq.~\eqref{Htotsuc}] to determine the transition rate from the initial state $\ket{\psi_i} = \ket{\psi_{j,+},n_i}$ to the final state $\ket{\psi_f} = \ket{\psi_{j,-},n_f}$, where $n_i$ ($n_f$) is the initial (final) occupancy of the phonon state and $\psi_{j,+}$ ($\psi_{j,-}$) is the wavefunction corresponding to the TLS in its excited (ground) state. Enforcing $n_f = n_i + 1$, as well as $E_j = \hbar \omega_q$ (when the TLS de-excites, it creates a single phonon of frequency $\omega_q$), while averaging over the initial phonon states and summing over the final phonon states, gives the total TLS de-excitation rate \cite{phillips_1987, enss_2005}
\begin{equation}
\upsilon_{j,-} = \left( \frac{\Delta_{0j}}{E_j} \right)^2 \sum_q \frac{\pi} {\omega_q} \left( \braket{n_q} + 1 \right) \left| \tensor{\gamma_j}:\tensor{\varepsilon}_q(\vec{r}_j) \right|^2 \delta(E_j - \hbar \omega_q),
\label{Fermiprob}
\end{equation}
where $\braket{n_q} = ( e^{\hbar \omega_q / k_{\rm B} T} - 1)^{-1}$ is the average phonon occupation of the $q$th mechanical mode according to Bose-Einstein statistics. Inputting this expression into Eq.~\eqref{relaxrate}, the TLS relaxation rate is then be found to be \cite{behunin_2016}
\begin{equation}
\tau_j^{-1} = \left( \frac{\Delta_{0j}}{E_j} \right)^2 \sum_q \frac{\pi} {\omega_q} \coth \left( \frac{E_j}{2 k_{\rm B} T} \right) \left| \tensor{\gamma_j}:\tensor{\varepsilon}_q(\vec{r}_j) \right|^2 \delta(E_j - \hbar \omega_q).
\label{finrelaxrate}
\end{equation}

To analyze how the delay in equilibration due this finite relaxation rate affects the dissipation of acoustic energy in each of the mechanical modes, we again look to the Hamiltonian in Eq.~\eqref{Htotsuc} to determine the Heisenberg equation of motion for $\hat{b}_q$ as
\begin{equation}
\begin{split}
\dot{\hat{b}}_q &= \frac{i}{\hbar} [ \hat{H}, \hat{b}_q ] \\
&= - \left( i \omega_q + \frac{\Gamma_{q,{\rm n}}}{2} \right) \hat{b}_q - \frac{i}{\hbar} \sum_j \left( \mu_{qj} \hat{\sigma}_x + \nu_{qj} \hat{\sigma}_z \right) - \sqrt{\Gamma_{q,{\rm n}}} \hat{b}_{q, {\rm n}},
\label{bqheis}
\end{split}
\end{equation}
where we have used the fact that $\frac{i}{\hbar} [ \hat{H}_\Gamma , \hat{b}_q ] = - \frac{\Gamma_{q,{\rm n}}}{2} \hat{b}_q - \sqrt{\Gamma_{q,{\rm n}}} \hat{b}_{q, {\rm n}}$, where $\Gamma_{q,{\rm n}}$ is the damping rate for the $q$th mechanical mode due to sources other than the TLS ensemble and $\hat{b}_{q, {\rm n}}$ is a drive term due to noise (both thermal and quantum) leaking in from the environment \cite{hauer_2015, clerk_2010}. Taking the expectation value of Eq.~\eqref{bqheis}, we find an analogous equation of motion for $\beta_q$ as
\begin{equation}
\dot{\beta}_q = - \left( i \omega_q + \frac{\Gamma_{q,{\rm n}}}{2} \right) \beta_q - \frac{i}{\hbar} \sum_j \nu_{qj} s_j - \sqrt{\Gamma_{q,{\rm n}}} \beta_{q, {\rm n}},
\label{betaqheis}
\end{equation}
where we have neglected the term proportional to $\mu_{qj}$. Fourier transforming Eq.~\eqref{betaqheis} and grouping terms proportional to $\beta_q$, while using the fact that only the dynamical part of $s_j$ [{\it i.e.}, $\delta s_j$ -- see Eq.~\eqref{TLSdevdecayfreq}] will contribute to the mechanical damping, we find the expression for the total dissipation rate of the $q$th mechanical mode as
\begin{equation}
\Gamma_{q} = \Gamma_{q,{\rm n}} + \Gamma_{q,{\rm TLS}},
\label{totdampq}
\end{equation}
where
\begin{equation}
\Gamma_{q,{\rm TLS}} = \sum_j \left( \frac{\Delta_j}{E_j} \right)^2 \frac{\left| \tensor{\gamma_j} : \tensor{\varepsilon}_q(\vec{r}_j) \right|^2}{k_{\rm B} T} {\rm sech}^2 \left( \frac{E_j}{2 k_{\rm B} T} \right) \frac{\tau_j}{1 + \omega_q^2 \tau_j^2},
\label{TLSdampq}
\end{equation}
is the mechanical damping rate due to the TLS-phonon relaxation interaction. We note that in the situation where TLS damping dominates ({\it i.e.}, $\Gamma_{q,{\rm TLS}} \gg \Gamma_{q,{\rm n}}$) for a given mode, we can take $\Gamma_q \approx \Gamma_{q,{\rm TLS}}$, as is done for the fits in Fig.~\ref{fig4}.

\subsection{Determination of $\left| \tensor{\gamma_j}:\tensor{\varepsilon}_q(\vec{r}_j) \right|^2$}
\label{dettens}

In general, the product $\tensor{\gamma_j}:\tensor{\varepsilon}_q(\vec{r}_j)$ found in Eq.~\eqref{TLSdampq} is a complicated, spatially varying sum over a number of tensor components. However, by using the local symmetries of the simple cubic lattice of crystalline silicon, as well as making some assumptions about our TLS ensemble, we can simplify this quantity considerably. We begin by expressing the deformation potential tensor as $\tensor{\gamma}_j = \tensor{R} : \tensor{W_j} = R_{abcd} W_{j,ab}$ \cite{anghel_2007}, where $\tensor{R}$ is a 4th rank tensor that describes the TLS environment and 
\begin{equation}
\tensor{W_j} = \begin{bmatrix} w_{j,x}^2 & w_{j,x} w_{j,y} & w_{j,x} w_{j,z} \\ w_{j,x} w_{j,y} & w_{j,y}^2 & w_{j,y} w_{j,z} \\ w_{j,x} w_{j,z} & w_{j,y} w_{j,z} & w_{j,z}^2 \end{bmatrix}
\label{Wtens}
\end{equation}
is a 2nd rank tensor that characterizes the orientation of each TLS. Here, $w_{j,x} = \sin(\theta_j) \cos(\phi_j)$, $w_{j,y} = \sin(\theta_j) \sin(\phi_j)$ and $w_{j,z} = \cos(\theta_j)$ are the components of the unit vector parallel to the defect's elastic dipole moment, with $\theta_j$ and $\phi_j$ specifying its orientation \cite{anghel_2007,behunin_2016}. Using this formalism, the tensor product found in $\mu_{qj}$, $\nu_{qj}$ and $\Gamma_{q,{\rm TLS}}$ can then be written as $\tensor{\gamma_j}:\tensor{\varepsilon}_q(\vec{r}_j) = R_{abcd} W_{j,ab} \varepsilon_{q,cd}(\vec{r}_j)$. 

Due to the simple cubic symmetry of the silicon lattice, $\tensor{R}$ will have only three independent parameters \cite{anghel_2007}, namely $R_{xxxx} = R_{yyyy} = R_{zzzz} = R_{11}$, $R_{xxyy} = R_{xxzz} = R_{yyzz} = R_{yyxx} = R_{zzxx} = R_{zzyy} = R_{12}$ and $R_{xyxy} = R_{yxyx} = R_{xyyx} = R_{yxxy} = R_{xzxz} = R_{zxzx} = R_{xzzx} = R_{zxxz} = R_{yzyz} = R_{zyzy} = R_{yzzy} = R_{zyyz} = R_{44}$, directly analogous to the elasticity tensor of the system (see Appendix \ref{strain}). Furthermore, assuming the TLS ensemble is uniformly distributed (both in spatial density and orientation), we can average $\left| \tensor{\gamma_j}:\tensor{\varepsilon}_q(\vec{r}_j) \right|^2$ over the total volume $V$ of the resonator and the solid angle of TLS orientations, resulting in \cite{behunin_2016}
\begin{equation}
\braket{\left| \tensor{\gamma_j}:\tensor{\varepsilon}_q(\vec{r}_j) \right|^2}_V = \frac{3 \omega_q^2}{V \rho} \sum_\eta \frac{\gamma_\eta^2}{c_\eta^2} e_{q\eta}.
\label{tensprodave}
\end{equation}
Here, the sum is over the three different phonon polarizations (one longitudinal and two transverse), where $\gamma_\eta$, $c_\eta$ and $e_{q\eta}$ are the deformation potential, speed of sound and fraction of the resonance mode's energy associated with each polarization. In terms of the components of $\tensor{R}$, the deformation potentials for each phonon polarization are given by
\begin{equation}
\begin{split}
\gamma_l &= \sqrt{\frac{2 R_{11}^2 + 7 R_{12}^2 + 6 R_{11}R_{12} + 4 R_{44}^2}{45}}, \\
\gamma_{t_1} &= \sqrt{\frac{(R_{11}-R_{12})^2}{45}}, \\
\gamma_{t_2} &= \frac{2 R_{44}}{\sqrt{45}},
\label{defpots}
\end{split}
\end{equation}
while explicit forms of $c_\eta$ and $e_{q \eta}$ are given by Eqs.~\eqref{soundspeeds} and \eqref{polfracs} in Appendix \ref{strain}. 

\subsection{Coupling to Ensembles of TLS Defects with Varying Energy Distributions}
\label{TLSens}

Using the results previously obtained in this appendix, we are now equipped to determine the mechanical dissipation due to coupling to a given ensemble of TLS defects. Starting with the relaxation rate, we input the result for $ \braket {\left| \tensor{\gamma_j}:\tensor{\varepsilon}_q(\vec{r}_j) \right|^2}_V$ from Eq.~\eqref{tensprodave} into Eq.~\eqref{finrelaxrate} to obtain
\begin{equation}
\tau_j^{-1} = \frac{3 \pi}{V \rho \hbar} \sum_{q,\eta} \frac{\Delta_{0j}^2}{E_j} \frac{\gamma_\eta^2}{c_\eta^2} e_{q\eta} \coth \left( \frac{E_j}{2 k_{\rm B} T} \right) \delta(E_j - \hbar \omega_q).
\label{relaxratesum}
\end{equation}
To evaluate the sum over $q$, we must carefully consider the density of states, $D_\eta(\omega)$, associated with $\eta$ polarized phonons. For the system at hand, a discrete density of states associated with the mechanical modes of the resonator would seem to be an obvious choice. However, because a large number of these modes are thermally populated for the temperature range considered (at $T = $ 10 mK, modes with frequencies up to $\omega_q / 2 \pi = 144$ MHz have at least one phonon on average), we can instead use the simpler continuum (Debye) density of states \cite{behunin_2016}. That said, we still need to determine the dimensionality of this density of states. This is done by comparing the characteristic dimensions of the device to its shortest thermal phonon wavelength, $\lambda_{\rm th} = 2 \pi \hbar c_{\rm min}/k_{\rm B} T$, where $c_{\rm min}$ is the smallest speed of sound in the material. If any of these dimensions are smaller than $\lambda_{\rm th}$, then the device is considered to be dimensionally-reduced in that direction \cite{behunin_2016,seoanez_2008}. For silicon, we take $c_{\rm min} = c_{t_1} = 4679$ m/s, such that $\lambda_{\rm th} \approx 225~{\rm nm} \cdot {\rm K}/T$. Therefore, the resonator considered in this work, which has cross-sectional dimensions of $w =$ 200 nm and $d =$ 250 nm (cross-sectional area of $A = w \times d =$ 5.0$\times$10$^{-14}$ m$^2$), can be treated as one-dimensional for $T \lesssim 1$ K. In the temperature range $T > 1$ K, we assume the resonator to be quasi-one-dimensional, such that for all relevant temperatures we can use the one-dimensional phononic density of states, $D_\eta(\omega) = L / \pi c_\eta$, where $L$ is the length of the mechanical resonator. With this choice of density of states, we can replace the sum in Eq.~\eqref{relaxratesum} with $\displaystyle \sum_q \rightarrow \int D_\eta(\omega) d\omega$, which upon performing the integral, results in
\begin{equation}
\tau_j^{-1} = \frac{1}{A \rho \hbar^2} \sum_\eta \frac{\Delta_{0j}^2}{E_j} \frac{\gamma_\eta^2}{c_\eta^3}  \coth \left( \frac{E_j}{2 k_{\rm B} T} \right).
\label{sumrelaxrate}
\end{equation}
Here, we have assumed $e_{ql} = e_{qt_1} = e_{qt_2} \approx 1/3$, as this is the average value for each fraction when a large number of mechanical modes are considered. We further note that each phonon polarization will in general have a unique deformation potential, $\gamma_\eta$, however, determining exact values for these parameters is beyond the scope of this work. Therefore, we further simplify this expression for the relaxation rate by assuming $\gamma_l = \gamma_{t_1} = \gamma_{t_2} = \gamma$, as well as introducing an effective speed of sound $\displaystyle c_{\rm e} = \left( \sum_\eta \frac{1}{c_\eta^3} \right)^{-1/3}$ = 3965 m/s, such that
\begin{equation}
\tau^{-1} = \frac{\gamma^2}{A \rho c_{\rm e}^3 \hbar^2} \frac{\Delta_{0}^2}{E} \coth \left( \frac{E}{2 k_{\rm B} T} \right).
\label{apprelaxrate}
\end{equation}
Finally, inputting this relaxation rate, as well as the spatially averaged value of $\left| \tensor{\gamma_j}:\tensor{\varepsilon}_q(\vec{r}_j) \right|^2$ from Eq.~\eqref{tensprodave} into Eq.~\eqref{TLSdampq} and replacing the sum over the TLS ensemble with an integral over the TLS density of states, $\displaystyle \sum_j \rightarrow \int_0^\infty \int_0^\infty V h(\Delta,\Delta_0) d \Delta d \Delta_0$ \cite{hunklinger_1976}, we get the ensemble-averaged TLS-induced damping rate
\begin{equation}
\begin{gathered}
\Gamma_{\rm TLS} = \frac{\gamma^2}{\rho c_q^2 k_{\rm B} T} \int_0^\infty \int_0^\infty  \left( \frac{\Delta}{E} \right)^2 {\rm sech}^2 \left( \frac{E}{2 k_{\rm B} T} \right) \\
\times \frac{\omega_{\rm m}^2 \tau}{1 + \omega_{\rm m}^2 \tau^2} h(\Delta,\Delta_0) d \Delta d \Delta_0,
\label{TLSdampfin}
\end{gathered}
\end{equation}
where, like in Eq.~\eqref{apprelaxrate}, we have introduced a mode-dependent effective speed of sound $ \displaystyle c_q = \left( 3 \sum_\eta \frac{e_{q\eta}}{c_\eta^2} \right)^{-1/2}$. We note that in Eqs.~\eqref{apprelaxrate} and \eqref{TLSdampfin} we have taken $\omega_{\rm q} \rightarrow \omega_{\rm m}$, as well as dropped the explicit subscripts $j$ and $q$, to match the notation of Eqs.~\eqref{tauinv} and \eqref{GamTLS}.

The functional form of $h(\Delta,\Delta_0)$ is chosen to characterize the energy distribution of the TLS ensemble and, depending on the dimensionality of the system, can have a drastic effect on the temperature dependence of the TLS-induced mechanical damping rate \cite{behunin_2016,phillips_1987}. In the standard tunneling model, this energy density function has the form
\begin{equation}
h_{\rm a}(\Delta,\Delta_0) = \frac{P_0}{\Delta_0},
\label{hamo}
\end{equation}
where $P_0$ is a constant that characterizes the density of states of the TLS ensemble \cite{anderson_1972,phillips_1972,phillips_1987,esquinazi_1998}, typically on the order of 10$^{44}$ J$^{-1}$ m$^{-3}$ for glassy solids \cite{behunin_2016,seoanez_2008}. An energy density function of this form reflects the broad distribution of $\Delta$ and $\Delta_0$ exhibited for amorphous TLS distributions and for the one-dimensional resonator geometry considered here, results in the damping rate
\begin{equation}
\begin{gathered}
\Gamma_{\rm a} = \frac{\gamma^2 P_0}{\rho c_q^2 k_{\rm B} T} \int_0^\infty \int_0^\infty \frac{\Delta^2}{\Delta_0 E^2}~{\rm sech}^2 \left( \frac{E}{2 k_{\rm B} T} \right) \\ 
\times \frac{\omega_{\rm m}^2 \tau}{1 + \omega_{\rm m}^2 \tau^2} d \Delta d \Delta_0.
\label{TLSdampa}
\end{gathered}
\end{equation}
At low temperatures [$\omega_{\rm m} \tau_{\rm min} \gg 1$, where $\tau_{\rm min} = \tau(E = \Delta_0)$], this mechanical damping rate can be approximated as
\begin{equation}
\Gamma_{\rm a,LT} \approx \frac{\pi^2 \gamma^4 P_0 k_{\rm B} T}{6 A \rho^2 \hbar^2 c_q^2 c_{\rm e}^3},
\label{adampLT}
\end{equation}
which is linear in $T$ as expected \cite{behunin_2016}. Meanwhile, at high-temperatures ($\omega_{\rm m} \tau_{\rm min} \ll 1$) we find
\begin{equation}
\Gamma_{\rm a,HT} \approx \frac{\pi \gamma^2 P_0 \omega_{\rm m}}{2 \rho c_q^2}.
\label{adampHT}
\end{equation}
We note that while the temperature-dependence of $\Gamma_{\rm a,LT}$ differs significantly from the $T^3$-dependence seen in a number of bulk amorphous solids \cite{pohl_2002}, for high temperatures, the mechanical damping rate approaches the same constant value regardless of the dimensionality of the system \cite{behunin_2016}, minimizing the effect of our choice of a one-dimensional phonon density of states for $T >$ 1 K.

On the other hand, for TLS ensembles that exhibit crystalline behaviour, a narrower distribution in TLS energies exists. To account for this, Phillips \cite{phillips_1988} suggested a distribution function of the form
\begin{equation}
h_{\rm c}(\Delta,\Delta_0) = D_0 \sqrt{\frac{2}{\pi}} \frac{1}{\delta \Delta} e^{-\frac{1}{2} \left( \frac{\Delta}{\delta \Delta} \right)^2} \delta(\Delta_0 - \bar{\Delta}_0),
\label{hcryst}
\end{equation}
that is, the crystalline nature of the TLS ensemble results in a well-defined tunneling energy of $\bar{\Delta}_0$ and a gaussian spread in the asymmetry energy, with a standard deviation of $\delta \Delta$ centered around $\Delta = 0$. We note that with this choice of distribution function, we need only consider the relevant case of $\omega_{\rm m} \tau \gg 1$, due to the fact that the experimentally measured dissipation increases monotonically with temperature for each of the mechanical modes studied in this paper (see in Fig.~\ref{fig4}) \cite{phillips_1988}. This allows us to approximate the mechanical damping rate in Eq.~\eqref{TLSdampfin} as
\begin{equation}
\begin{gathered}
\Gamma_{\rm c} = \frac{ \gamma^2 D_0}{\rho c_q^2 k_{\rm B} T \delta \Delta} \sqrt{\frac{2}{\pi}} \int_0^\infty \left( \frac{\Delta}{\bar{E}} \right)^2 {\rm sech}^2 \left( \frac{\bar{E}}{2 k_{\rm B} T} \right) \\
\times e^{-\frac{1}{2} \left( \frac{\Delta}{\delta \Delta} \right)^2} \tau^{-1}(\Delta, \bar{\Delta}_0) d \Delta,
\label{TLSdampc}
\end{gathered}
\end{equation}
where $\bar{E}^2 = \Delta^2 + \bar{\Delta}_0^2$ and we have expressed $\tau$ as an explicit function of $\Delta$ and $\Delta_0$. To examine the low temperature limit of the damping due to this crystalline TLS distribution, we take $T \ll (\delta \Delta^2 + \bar{\Delta}_0^2)^{1/2} / k_{\rm B}$, allowing for the approximations ${\rm sech}^2 \left( \bar{E} / 2 k_{\rm B} T \right) \approx 4 e^{-\bar{E}/k_{\rm B} T}$ and $\coth \left( \bar{E} / 2 k_{\rm B} T \right) \approx 1$ over the regions of integration in Eq.~\eqref{TLSdampc} that provide the majority of the contribution to $\Gamma_{\rm c}$, resulting in a low-temperature dependence according to
\begin{equation}
\Gamma_{\rm c,LT} \approx \frac{ 4 \gamma^4 D_0 \bar{\Delta}_0^2}{A \rho^2 \hbar^2 c_q^2 c_{\rm e}^3 k_{\rm B} T \delta \Delta} \sqrt{\frac{2}{\pi}} \int_0^\infty \frac{\Delta^2}{\bar{E}^3} e^{-\frac{\bar{E}}{k_{\rm B}T}} e^{-\frac{1}{2} \left( \frac{\Delta}{\delta \Delta} \right)^2} d \Delta.
\label{cdampLT}
\end{equation}
In the opposite limit of $T \gg (\delta \Delta^2 + \bar{\Delta}_0^2)^{1/2} / k_{\rm B}$, we have ${\rm sech}^2 \left( \bar{E} / 2 k_{\rm B} T \right) \approx 1$ and $\coth \left( \bar{E} / 2 k_{\rm B} T \right) \approx 2 k_{\rm B} T / \bar{E}$, such that the high temperature limit for $\Gamma_{\rm c}$ is given by
\begin{equation}
\Gamma_{\rm c,HT} \approx \frac{ 2 \gamma^4 D_0 \bar{\Delta}_0^2}{A \rho^2 \hbar^2 c_q^2 c_{\rm e}^3 \delta \Delta} \sqrt{\frac{2}{\pi}} \int_0^\infty \frac{\Delta^2}{\bar{E}^4} e^{-\frac{1}{2} \left( \frac{\Delta}{\delta \Delta} \right)^2} d \Delta.
\label{cdampHT}
\end{equation}
Here, the factor of $k_{\rm B} T$ from the approximation of $\coth \left( E / 2 k_{\rm B} T \right)$ cancels that in the denominator of Eq.~\eqref{TLSdampc}, such that $\Gamma_{\rm c,HT}$ is temperature independent, similar to the high-temperature limit of the damping rate due to the amorphous TLS distribution, albeit at a different value.

\section{Fits to Amorphous and Crystalline Two-Level System Damping Models}
\label{tlsfits}

\begin{figure*}[t!]
\centerline{\includegraphics[width=6in]{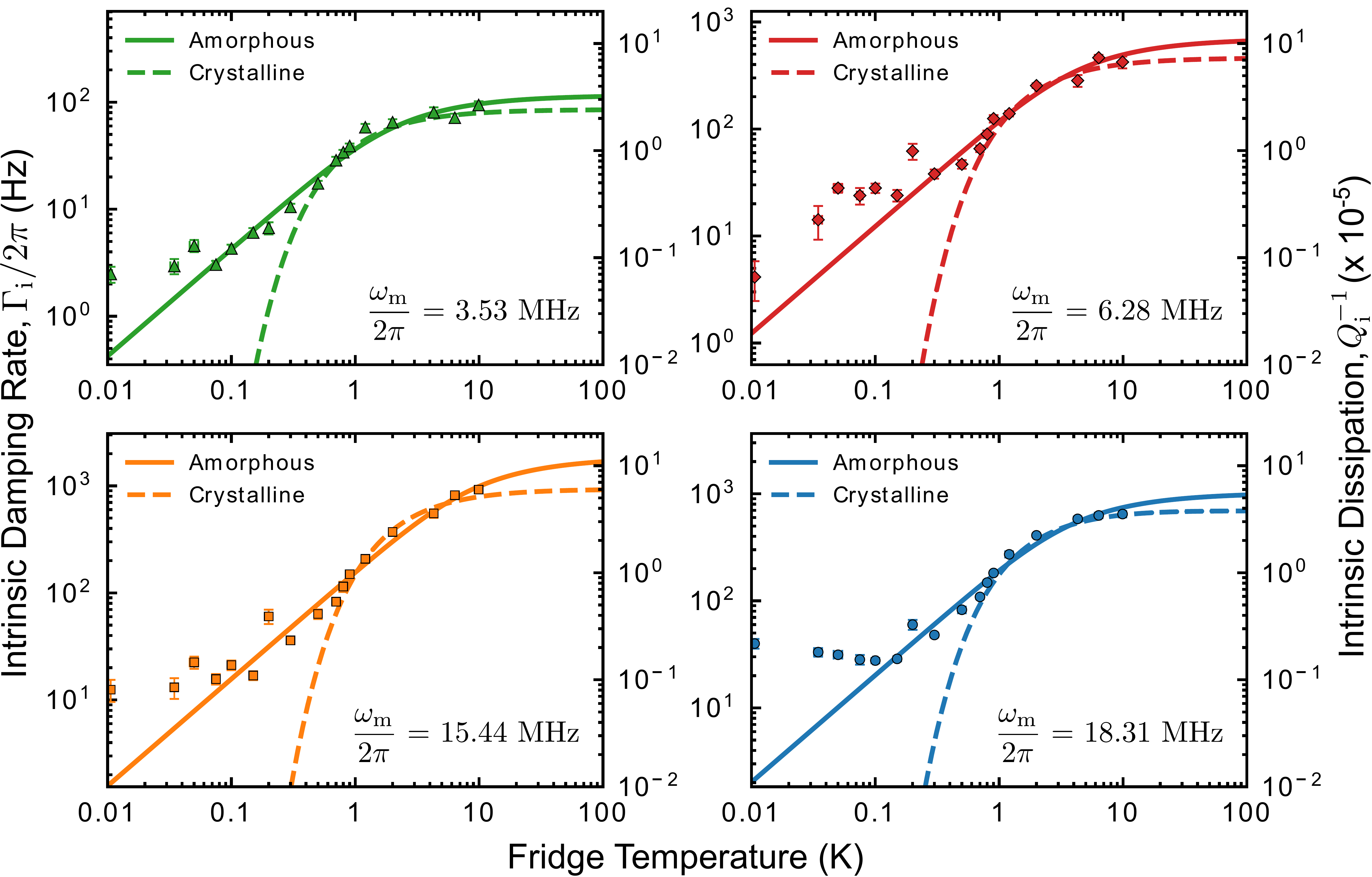}}
\caption{{\label{fig9}} The intrinsic damping rate, $\Gamma_{\rm i}$, for each of the four studied mechanical modes versus temperature, with the right axis displaying their intrinsic dissipation, $Q_{\rm i}^{-1} = \Gamma_{\rm i}/\omega_{\rm m}$. Each mode is fit to both the amorphous (solid line) and crystalline (dashed line) TLS damping models given by Eqs.~\eqref{TLSdampa} and \eqref{TLSdampc}, respectively. From these fits we extract $P_0$ and $\gamma$ for the amorphous TLS damping model, as well as $\bar{\Delta}_0$, $\delta \Delta$ and $\gamma^4 D_0$ for the crystalline TLS damping model for each mechanical mode. The parameters from each fit are displayed in Tables \ref{TLStaba} and \ref{TLStabc} for the amorphous and crystalline models, respectively.}
\end{figure*}

Here, we present fits of the one-dimensional TLS dissipation models, with both amorphous and crystalline distributions, to the mechanical damping rate data for the four mechanical modes studied in this work. These fits are displayed in Fig.~\ref{fig9}, with the parameters extracted from each displayed in Tables \ref{TLStaba} and \ref{TLStabc}. As one can see in Fig.~\ref{fig9}, the crystalline model exhibits a much more rapid decline in dissipation versus temperature as compared to the amorphous model, far undershooting the measured damping rates for $T \lesssim$ 500 mK. Furthermore, this crystalline model plateaus to a constant value at high temperature that is slightly smaller than the amorphous model predicts. The amorphous TLS damping model is therefore a better fit to our data, implying that the low-temperature dissipation in our mechanical modes is caused by coupling to a glassy distribution of defects.

\begin{table}[h!]
\begin{tabular}{ ccc }
\hline
$\omega_{\rm m}/2 \pi$ & $P_0$ & $\gamma$ \\  
 ~(MHz)~ & ~(J$^{-1}$ m$^{-3}$)~ & ~(eV)~ \\ \hline
\hline
3.53 & (9.7 $\pm$ 3.4)$\times 10^{43}$ & 1.3 $\pm$ 0.1 \\

6.28 & (4.0 $\pm$ 2.7)$\times 10^{44}$ & 1.2 $\pm$ 0.2 \\

15.44 & (3.6 $\pm$ 1.9)$\times 10^{44}$ & 1.3 $\pm$ 0.2 \\

18.31 & (7.0 $\pm$ 4.3)$\times 10^{43}$ & 2.2 $\pm$ 0.4 \\

\hline
\end{tabular}
\caption{Summary of the TLS density of states parameter, $P_0$, and deformation potential, $\gamma$, for the fits to the mechanical damping rate using the amorphous TLS damping model [{\it i.e.}, Eq.~\eqref{TLSdampa}] in Fig.~\ref{fig9} (solid line), as well as Fig.~\ref{fig4}. The uncertainty in each parameter is given by their standard deviations from the fit.}
\label{TLStaba}
\end{table}

\begin{table}[h!]
\begin{tabular}{ ccccc }
\hline
$\omega_{\rm m}/2 \pi$ & $\bar{\Delta}_0$ & $\delta \Delta$ & $\gamma^4 D_0$ \\  
 (MHz) & ($\mu$eV) & (meV) & (eV$^4$ m$^{-3}$) \\ \hline
\hline
3.53 & 74 $\pm$ 6 & 24.6 $\pm$ 0.7 & (3.6 $\pm$ 0.1) $\times 10^{24}$ \\

6.28 & 133 $\pm$ 19 & 37.3 $\pm$ 1.1 & (1.7 $\pm$ 0.2) $\times 10^{25}$ \\

15.44 & 165 $\pm$ 12 & 19.3 $\pm$ 397 & (2 $\pm$ 30) $\times 10^{25}$ \\

18.31 & 129 $\pm$ 8 & 3.5 $\pm$ 2.0 & (3.0 $\pm$ 1.7) $\times 10^{24}$ \\

\hline
\end{tabular}
\caption{Summary of the tunneling energy, $\bar{\Delta}_0$, and spread in asymmetry energy, $\delta \Delta$, as well as $\gamma^4 D_0$, for the fits to the mechanical damping rate using the crystalline TLS damping model [{\it i.e.}, Eq.~\eqref{TLSdampc}] in Fig.~\ref{fig9} (dashed line). The uncertainty in each parameter is given by their standard deviations from the fit.}
\label{TLStabc}
\end{table}

\section{Mechanical Resonator Heating Model}
\label{heatmodel}

\begin{figure}[t!]
\centerline{\includegraphics[width=3in]{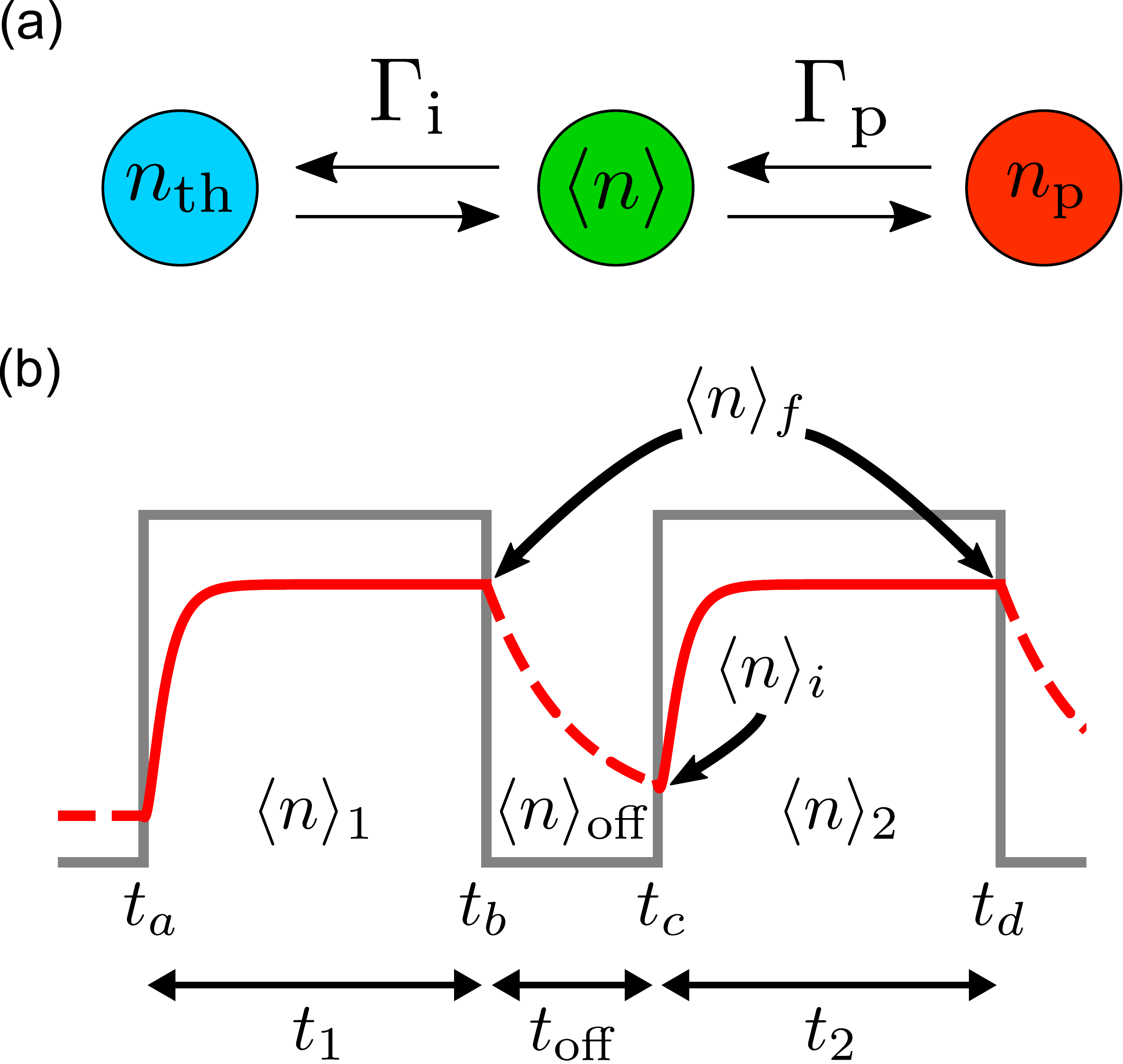}}
\caption{{\label{fig10}} (a) Block diagram of the heating model for the experimentally relevant case where the mechanical mode is coupled to a cold environmental bath at its intrinsic damping rate, $\Gamma_{\rm i}$, and a hot photon-induced bath at a rate, $\Gamma_{\rm p}$, each with phonon occupancies of $n_{\rm th}$ and $n_{\rm p}$, respectively. (b) Diagram of the double pulse measurement scheme used in this work. The solid grey line indicates the duty cycle of the laser, while the solid (dashed) red line expresses the phonon occupation of the mechanical mode with the laser on (off).}
\end{figure}

We model the thermalization of a given mode of our mechanical resonator as a harmonic oscillator at frequency, $\omega_{\rm m}$, coupled at its intrinsic damping rate, $\Gamma_{\rm i}$, to the device's cold environmental bath at temperature, $T_{\rm b}$, as well as at a rate, $\Gamma_{\rm p}$, to a hot phonon bath at temperature, $T_{\rm p}$, created by optical absorption of measurement photons and/or radiation pressure backaction, as depicted schematically in Fig.~\ref{fig10}(a). Due to the high quality factors of the mechanical modes considered in this paper, we can also treat both of the environmental and photon-induced baths as harmonic oscillators at the mechanical frequency, each with an average occupancy of $n_{\rm th}$ and $n_{\rm p}$, respectively. In this situation, the rate equation for the average occupation of the mechanical mode will be given by \cite{meenehan_2015}
\begin{equation}
\dot{\braket{n}} = - \Gamma \braket{n} + \Gamma_{\rm i} n_{\rm th} + \Gamma_{\rm p} n_{\rm p},
\label{heatrate}
\end{equation}
with $\Gamma = \Gamma_{\rm i} + \Gamma_{\rm p}$ being the total rate at which the mode equilibrates to the two baths. We note that our treatment of the mechanical mode occupation dynamics differs from that of Ref.~\cite{meenehan_2015}, as we have not included a time-dependent term proportional to $n_{\rm p}$ that accounts for the finite equilibration time of the hot photon-induced bath. This is justified by the fact that the thermal relaxation time for our device, found by approximating each half of our resonator as a simple rectangular beam 10 $\mu$m in length \cite{zener_1948, lifshitz_2000} with a cross-section-limited thermal conductivity \cite{heron_2009, heron_2010, cahill_2014}, is roughly 20 ns. Therefore, our measurement scheme, with a temporal resolution on the order of 1 $\mu$s, is unable to resolve this thermalization process and we neglect to include this time-dependent term in Eq.~\eqref{heatrate}. 

Solving the rate equation given in \eqref{heatrate}, we find the time-dependent mechanical mode occupancy to be
\begin{equation}
\braket{n}(t) = \braket{n}(t_0)e^{-\Gamma (t-t_0)} + n_{\rm eq} \left(1 - e^{-\Gamma (t-t_0)} \right),
\label{heatsol}
\end{equation}
where $\braket{n}(t_0)$ is the phonon occupancy at the initial time $t_0$ and $n_{\rm eq} = (\Gamma_{\rm i} n_{\rm th} + \Gamma_{\rm p} n_{\rm p})/\Gamma$ is the occupancy of the mechanical mode at times $t \gg \Gamma^{-1}$, long enough that the mode is able to equilibrate to an average of the bath occupations, weighted by their coupling rates. Furthermore, if the connection to the hot photon-induced bath is severed ({\it i.e.}, by turning the laser off), we take $\Gamma_{\rm p}=0$ such that the mechanical mode occupation will tend towards equilibrium with the environmental bath at its intrinsic damping rate  according to
\begin{equation}
\braket{n}(t) = \braket{n}(t_0)e^{-\Gamma_{\rm i} (t-t_0)} + n_{\rm th} \left(1 - e^{-\Gamma_{\rm i} (t-t_0)} \right).
\label{coolsol}
\end{equation}

For the experiment considered in this paper, we measure the low temperature damping rate of our mechanical device, using the pump/probe measurement outlined in Appendix \ref{expdet}. This procedure can be described by the general two pulse scheme depicted in Fig.~\ref{fig10}(b), where a pump pulse that turns on at $t=t_a$ and off at $t=t_b$ (pulse length $t_1 = t_b - t_a$), is followed by a probe pulse that turns on at $t=t_c$ and off at $t=t_d$ (pulse length $t_2 = t_d - t_c$), with a delay between the two pulses of $t_{\rm off} = t_c - t_b$. For this situation, the occupation of the mechanical mode during the pump pulse will evolve in time according to Eq.~\eqref{heatsol} as
\begin{equation}
\braket{n}_1(t) = \braket{n}(t_a)e^{-\Gamma (t-t_a)} + n_{\rm eq} \left(1 - e^{-\Gamma (t-t_a)} \right).
\label{pumpsol}
\end{equation}
Once the pump pulse has been turned off, the resonator's occupancy will cool towards that of the environmental bath, as governed by Eq.~\eqref{coolsol} to give
\begin{equation}
\braket{n}_{\rm off}(t) = \braket{n}(t_b)e^{-\Gamma_{\rm i} (t-t_b)} + n_{\rm th} \left(1 - e^{-\Gamma_{\rm i} (t-t_b)} \right).
\label{offsol}
\end{equation}
Finally, the occupation of the mechanical mode during the probe pulse will obey
\begin{equation}
\braket{n}_2(t) = \braket{n}(t_c)e^{-\Gamma (t-t_c)} + n_{\rm eq} \left(1 - e^{-\Gamma (t-t_c)} \right).
\label{probesol}
\end{equation}
Assuming the experimentally relevant case of $t_1 = t_2 \gg \Gamma^{-1}$, the final occupancy at the end of either the pump or probe pulse will be given by $\braket{n}(t_b) = \braket{n}(t_d) = n_{\rm eq}$, while the initial occupancy of the mode at the beginning of the probe pulse can be found to be $\braket{n}(t_c) = n_{\rm eq} e^{-\Gamma_{\rm i} t_{\rm off}} + n_{\rm th} \left(1 - e^{-\Gamma_{\rm i} t_{\rm off}} \right)$. Using these two expressions, we can determine the ratio of the {\it measured} occupancy at the beginning of the probe pulse, $\braket{n}_i$, to the final {\it measured} occupancy of either the probe or the pump pulse, $\braket{n}_f$, as
\begin{equation}
\begin{split}
\frac{\braket{n}_i}{\braket{n}_f} &= \frac{\braket{n}(t_c) + n_{\rm imp}}{\braket{n}(t_b) + n_{\rm imp}} = \frac{\braket{n}(t_c) + n_{\rm imp}}{\braket{n}(t_d) + n_{\rm imp}} \\
&= \frac{\left(n_{\rm eq} - n_{\rm th} \right) e^{-\Gamma_{\rm i} t_{\rm off}} + n_{\rm th} + n_{\rm imp}}{n_{\rm eq} + n_{\rm imp}},
\label{ninfrat}
\end{split}
\end{equation}
where we have included the noise due to the imprecision of the measurement as an apparent phonon occupancy $n_{\rm imp}$. Using this equation, thermal ringdown data for the mechanical mode can be fit to extract its intrinsic damping rate, as is done in Fig.~\ref{fig3}(b).

We conclude on the note that it is often the case that $n_{\rm eq} \approx n_{\rm p} \gg n_{\rm th},n_{\rm imp}$, such that Eq.~\eqref{ninfrat} simplifies to
\begin{equation}
\frac{\braket{n}_i}{\braket{n}_f} \approx e^{-\Gamma_{\rm i} t_{\rm off}} + \frac{n_{\rm th} + n_{\rm imp}}{n_{\rm eq}},
\label{ninfratapp}
\end{equation}
as can be seen by the fact that $\braket{n}_i / \braket{n}_f \approx 1$ for $t_{\rm off} \ll \Gamma_{\rm i}^{-1}$ in Fig.~\ref{fig3}(b).

\end{appendix}

\end{document}